\documentclass{article}
\usepackage{jcappub}

\usepackage{epstopdf}
\usepackage{graphicx}
\usepackage{color}
\usepackage[usenames,dvipsnames]{xcolor}
\usepackage{hyperref}
\usepackage[export]{adjustbox}

\newcommand{\be}{\begin{equation}}
\newcommand{\ee}{\end{equation}}
\newcommand{\bea}{\begin{eqnarray}}
\newcommand{\eea}{\end{eqnarray}}
\newcommand{\bdm}{\begin{displaymath}}
\newcommand{\edm}{\end{displaymath}}

\newcommand{\Om}{\Omega}

\newcommand{\cub}{$^3$}

\newcommand{\lcdm}{$\Lambda$CDM}
\newcommand{\lcdmn}{$\nu$\lcdm}
\newcommand{\omegam}{$\Omega_{\rm m}$}

\newcommand{\smnu}{$\Sigma \, m_\nu$}
\newcommand{\smnut}{$\Sigma \, m_\nu=0,\,0.17,\,0.3,\,0.53$ eV}

\newcommand{\subfind}{\textsc{subfind}}

\newcommand{\msun}{$M_{\odot}$} 

\newcommand{\rfiveh}{$R_{200}$}

\newcommand{\hmone}{$\,h^{-1}$}

\newcommand{\planck}{{\it Planck}}

\def\Ms{\, h^{-1} \, M_\odot}
\def\Mpc{\, h^{-1} \, {\rm Mpc}}

\def\cGpc{\, h^{-3} \, {\rm Gpc}^3}
\def\kMpc{\, h \, {\rm Mpc}^{-1}}

\def\sigem{\sigma_{8,mm}}
\def\sigec{\sigma_{8,cc}}

\newcommand{\eq}[1]{Eq.~(\ref{#1})}
\newcommand{\eqs}[1]{Eqs.~(\ref{#1})}
\newcommand{\fig}[1]{Figure~\ref{#1}}
\newcommand{\figs}[1]{Figures~\ref{#1}}

\def\ie{{\em i.e.}~}
\def\eg{{{\em e.g.}}}

\def\halofit{\textsc{halofit}}

\newcommand{\gadget}{\textsc{gadget}}
\newcommand{\gadgetthree}{\textsc{gadget-3}}

\title{DEMNUni: The clustering of large-scale structures in the presence of massive neutrinos}

\author[a,b]{Emanuele Castorina,}
\author[c,d]{Carmelita Carbone,}
\author[c]{Julien Bel,}
\author[c,e]{Emiliano Sefusatti,}
\author[f,g]{Klaus Dolag}

\affiliation[a]{SISSA- International School for Advanced Studies,
Via Bonomea 265, I-34136 Trieste -- Italy}
\affiliation[b]{INFN - sezione di Trieste, via Valerio 2, I-34127 Trieste – Italy}
\affiliation[c]{INAF - Osservatorio Astronomico di Brera, via E. Bianchi 46, I-23807 Merate (LC) -- Italy}
\affiliation[d]{INFN - Sezione di Bologna, viale Berti Pichat 6/2, I-40127, Bologna (BO) -- Italy}
\affiliation[e]{INFN - Sezione di Padova, via Marzolo 8, I-35131 Padova -- Italy}
\affiliation[f]{Department of Physics, Ludwig-Maximilians-Universit\"{a}t, Scheinerstr. 1, D-81679 M\"{u}nchen, Germany}
\affiliation[g]{Max-Planck-Institut f\"{u}r Astrophysik, Karl-Schwarzschild Strasse 1, D-85748 Garching bei M\"{u}nchen, Germany}

\emailAdd{ecastori@sissa.it}
\emailAdd{carmelita.carbone@brera.inaf.it}
\emailAdd{julien.bel@brera.inaf.it}
\emailAdd{emiliano.sefusatti@brera.inaf.it}
\emailAdd{dolag@usm.uni-muenchen.de}

\abstract{We analyse the clustering features of Large Scale Structures (LSS) in the presence of massive neutrinos, employing a set of large-volume, high-resolution cosmological N-body simulations,  where neutrinos are treated as a separate collisionless fluid. The volume of 8$\cGpc$, combined with a resolution of about $8\times 10^{10}\Ms$ for the cold dark matter (CDM) component, represents a significant improvement over previous N-body simulations in massive neutrino cosmologies. In this work we focus, in the first place, on the analysis of nonlinear effects in CDM and neutrinos perturbations contributing to the total matter power spectrum. 
We show that most of the nonlinear evolution is generated exclusively by the CDM component. We therefore compare mildly nonlinear predictions from Eulerian Perturbation Theory (PT), and fully nonlinear prescriptions (\halofit) with the measurements obtained from the simulations. We find that accounting only for the nonlinear evolution  of the CDM power spectrum allows to recover the total matter power spectrum with the same accuracy as the massless case. Indeed, we show that,  the most recent version of the \halofit\ formula calibrated on $\Lambda$CDM simulations can be applied directly to the linear CDM power spectrum without requiring additional fitting parameters in the massive case. As a second step, we study the abundance and clustering properties of CDM halos, confirming that, in massive neutrino cosmologies, the proper definition of the halo bias should be made with respect to the {\em cold} rather than the {\em total} matter distribution, as recently shown in the literature. Here we extend these results to the redshift space, finding that, when accounting for massive neutrinos, an improper definition of the linear bias can lead to a systematic error of about 1-$2 \%$ in the determination of the linear growth rate from anisotropic clustering. This result is quite important if we consider that future spectroscopic galaxy surveys, as \eg\ Euclid, are expected to measure the linear growth-rate with statistical errors less than about $3 \%$ at $z\lesssim1$.}


\begin{document}

\maketitle

\section{Introduction}
\label{intro}

In the $\Lambda$CDM cosmological model the three active neutrinos of the standard model of particle physics are assumed to be massless. Nevertheless, already in 1998 the Super-Kamiokande collaboration presented evidence of neutrino oscillations \cite{FukudaEtal1998}, indicating that at least two neutrinos are massive, and, more recently, new neutrino oscillation experiments seem to exclude a vanishing flavor mixing angle at more than 10 $\sigma$ (see, {\eg} \cite{FogliEtal2012, ForeroTortoraValle2012}). Indeed, the study of the effects of neutrino masses on cosmological observables is of particular relevance for two, distinct, reasons. First, the absolute neutrino mass scale remains unknown and, in this respect, cosmology plays a key role in its determination, being gravity sensitive to the total neutrino mass, \smnu,  rather than to the mass splitting. Second, an accurate description of massive neutrino effects on LSS is required to avoid systematic errors in the determination of cosmological parameters, as the dark energy density and equation of state, whose measurements represent one of the main goals of current and future cosmological experiments.

Massive neutrino cosmologies have been extensively studied in the literature (see \cite{LesgourguesPastor2006, Lesgourgues2013} for a review). In particular the linear perturbation theory in the presence of massive neutrinos is well understood and it is widely used to derive constraints on \smnu\ and other cosmological parameters, from present and future CMB and galaxy surveys observations \cite{Hannestad2003, ReidEtal2010, ThomasAbdallaLahav2010, SwansonPercivalLahav2010, SaitoTakadaTaruya2011, Carbone_etal2011, DePutterEtal2012B, XiaEtal2012, RiemerSorensenEtal2012, ZhaoEtal2013, MoreEtal2013, GiusarmaEtal2013, WymanEtal2013, BattyeMoss2014, RiemerSorensenParkinsonDavis2014, BeutlerEtal2014, AbazajianEtal2015, BohringerChon2015}.  However, the increasing precision of cosmological parameter measurements further requires an accurate description of nonlinear corrections, which can be obtained by means of a direct analysis of the output of N-body simulations accounting for  a massive neutrino component. This kind of simulations turns to be computationally very expensive and quite challenging if  a good mass resolution and a large box size are required at the same time in order to build realistic mock catalogues for present and future galaxy and weak lensing surveys. 

The ``Dark Energy and Massive Neutrino Universe'' (DEMNUni) simulation project presented in this and in the companion paper \cite{CarbonePetkovaDolag2015}, addresses in a consistent way the problem of structure formation in massive neutrino cosmologies, and represents the state of the art of neutrino simulations in terms of volume and mass resolution. At present, as explained in more details in \S\ref{sec:sim} below, these simulations include a baseline $\Lambda$CDM model and three cosmologies characterised by different values of \smnu, with all the cosmologies sharing the same total matter density $\Omega_m$, as well as the same amplitude of primordial curvature perturbations. In the near future, the DEMNUni set will be extended with the inclusion of an evolving dark energy background, with different equations of state $w$, in order to study  the degeneracy between the neutrino mass \smnu\ and $w$  at the nonlinear level.

This work focuses on the analysis of the clustering features of LSS in massive neutrino cosmologies, as extracted from the DEMNUni simulation set, including, in the first place, an accurate comparison of the nonlinear total matter power spectrum with the most recent predictions in PT, over the range of scales where such predictions are expected to significantly improve over linear theory. To this end we will separately consider the clustering of the CDM and neutrino components as well as their cross-correlation, following previous indications that such distinctions are quite relevant for an accurate description of gravitational instability in these cosmological models (see, {\eg} \cite{IchikiTakada2012, CastorinaEtal2014, BlasEtal2014}).  At the same time, we will test different implementations of popular fitting formulae such as {\halofit} \cite{Smith_etal2003, Bird_etal2012, TakahashiEtal2012} describing the nonlinear regime. A specific study of non-Gaussian aspects of the matter distribution in massive neutrinos cosmologies is left to future work \cite{RuggieriEtal2015}. 

In addition, we will revise previous results on the abundance and clustering of dark matter halos, taking advantage of the large volume and mass resolution provided by the DEMNUni simulations. In this respect, our results provide further evidence that a description in terms of cold (rather than {\em total}) dark matter perturbations is indeed required for the extension of common fitting functions for the halo mass function to the case of massive neutrinos. 
In this respect, despite the significant systematic effect that the wrong assumption would lead to the interpretation of cluster abundance observations \cite{CostanziEtal2014}, these early indications have been so far ignored in the analysis of relevant data-sets (\eg\ \cite{Planck2015szhalos}). We believe that, whether or not particle-based neutrino simulations provide an faithful description of reality, such systematic uncertainties should be at the very least accounted for in any exhaustive error budget. Similar considerations can be extended to the interpretation of galaxy power spectrum measurements where the assumption of a constant linear bias with respect to the CDM or total matter power spectrum can lead to quite different constraints on the total neutrino mass \cite{CastorinaEtal2014}. In this work, for the first time we extend from real to redshift space our results on halo clustering in the presence of massive neutrinos, providing a first insight into possible, additional systematic effects due to erroneous assumptions in the modelling of galaxy clustering anisotropies in massive neutrino scenarios.

This paper is organised as follow. In \S\ref{sec:linear_nu} we give a brief description of linear perturbation theory in the presence of massive neutrinos. In \S\ref{sec:sim} we present the N-body runs and the details of the numerical setup. In \S\ref{sec:matterps} we present our measurements of all the components to the matter power spectrum, along with comparison with PT results and popular fitting formulae. In \S\ref{sec:mf} and \S\ref{sec:halops}, we focus our attention on the mass function and the power spectrum of halos both in real and redshift space. Finally, in \S\ref{sec:conclusions} we summarise the main results of the present work and draw our conclusions, highlighting several applications of the DEMNUni simulations to the testing of different observables and probes in the context of massive neutrinos cosmologies.

\section{Massive neutrinos perturbations}
\label{sec:linear_nu}

Massive neutrinos are considered hot dark matter, since they decouple in the early Universe as relativistic particles, just before the onset of Big Bang Nucleosynthesis, and become non-relativistic only at later times. This transition happens when the neutrino temperature drops below the value of their mass at redshift \cite{Takada_etal_2006}
\be
1+z_{nr} \simeq 1890 \left(\frac{m_{\nu}}{1 \,{\rm eV}}\right)\,.
\label{eq:z_nr}
\ee
After this time they contribute to the total energy density of the Universe as dark matter so that the total matter energy density is given by
\be
\Omega_m = \Omega_{cdm} + \Omega_{b} + \Omega_{\nu}\,,
\label{eq:Om}
\ee
where, in addition to the CDM and baryon components, $\Omega_\nu$ accounts for the energy density associated to the {\em massive} neutrino component. Since we will consider scales much larger than the Jeans length of baryons, in what follows we will define $\Omega_c\equiv\Omega_{cdm}+\Omega_b$ as the density corresponding to the sum of CDM and baryon densities, and, for simplicity, will generically refer to it as a single ``CDM'' contribution, denoted by the subscript ``{\em c}''. It can be shown that $\Omega_\nu$ is proportional to the total neutrino mass as
\be
\Omega_\nu =\frac{\sum m_{\nu}}{93.14\, h^2\, {\rm eV}}\,,
\ee
where the proportionality factor depends on the assumed photon temperature and neutrino to photon temperature ratio and it should be evaluated numerically in the most general case \cite{ManganoEtal2005}. As a result, at late times, \ie $z\ll z_{nr}$, the effect of neutrinos on the expansion rate of the Universe is completely degenerate with a change of the CDM and baryon components.

On the other hand, at the perturbation level massive neutrinos have a peculiar effect on matter density fluctuations. Defining the density contrasts for neutrino and CDM respectively as $\delta_\nu=\delta\rho_\nu/\bar{\rho}_\nu$ and $\delta_c=\delta\rho_c/\bar{\rho}_c$, where the total mass density is $\rho_m=\bar{\rho}_m+ \delta\rho_c+\delta\rho_\nu$, $\bar{\rho}_m=\bar{\rho}_c+\bar{\rho}_\nu$ being the total background matter density, we can write 
\be
\delta_m = (1-f_{\nu})\, \delta_{c} + f_{\nu}\, \delta_{\nu}\,,
\label{eq:dm}
\ee 
where $f_\nu$ represents the neutrino fraction defined as $f_\nu\equiv \Omega_{\nu} / \Omega_m$.  It follows that the total matter power spectrum can be written as the sum of three contributions corresponding respectively to the CDM power spectrum, $P_{cc}$, the neutrinos power spectrum, $P_{\nu\nu}$,  and the cross-power spectrum between CDM and neutrinos, $P_{c\nu}\langle\delta_c\delta_\nu^*\rangle$, that is
\be
P_{mm}= (1-f_{\nu})^2\, P_{cc}+ 2 f_{\nu} \,(1-f_{\nu})\,P_{c\nu} + f_{\nu}^2\, P_{\nu \nu}\, , 
\label{eq:Pmm}
\ee
which shows that the neutrino fraction has a direct impact on the total matter power spectrum by modifying the relative contributions of the two components. Eqs.~(\ref{eq:Om}) to (\ref{eq:Pmm}) introduce the notation adopted throughout the rest of the paper.

Over the age of the Universe, neutrinos travel an average distance that depends on their thermal velocity and, in turn, on their mass. This {\em free streaming length} determines the scale below which neutrinos density perturbations are washed-out, and is given by (see, \eg\,\cite{Lesgourgues2013})
\be
\lambda_{\rm FS} (m_\nu, z) \simeq 8.1 \frac{H_0\,(1+z)}{H(z)} \left(\frac{1\, {\rm eV} }{m_\nu} \right) \Mpc\,.
\label{eq:lfs}
\ee
\begin{figure}[t!]
\begin{center}
\includegraphics[width=.98\textwidth]{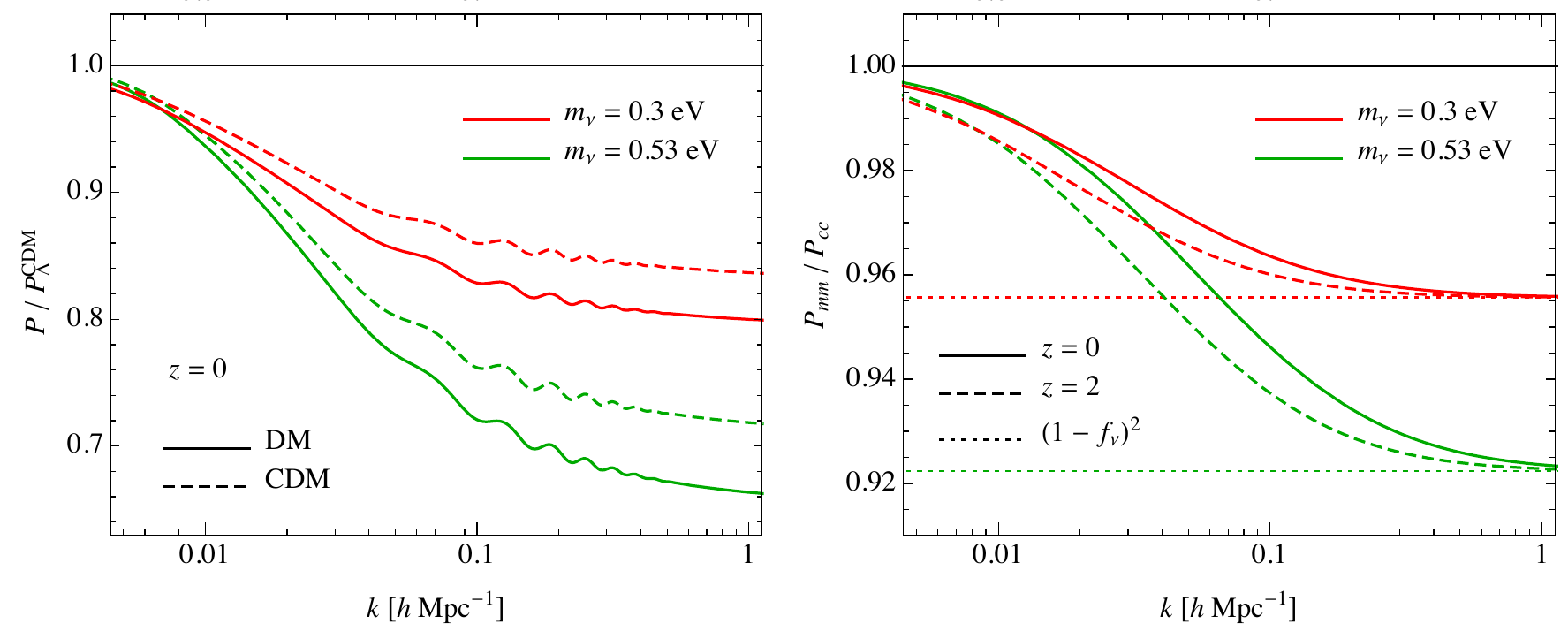}
\caption{\label{fig:PL} 
Linear theory results in massive neutrino cosmologies. {\em Left panel}: Ratio of the total matter power spectrum to the CDM power spectrum at redshifts $z=0$ ({\em continuous curves}) and $z=2$ ({\em dashed curves}) for two different values of the sum of neutrino masses, \smnu$=0.3$ eV in red and \smnu$=0.53$ eV in green. Dotted lines denote the asymptotic value at small scales of $(1-f_\nu)^2$. {\em Right panel}: ratio at $z=0$ of the total matter power spectrum ({\em continuous curves}) and CDM power spectrum ({\em dashed curves}) for the same two cosmologies to the $\Lambda$CDM prediction. }
\end{center}
\end{figure}

Notice that, for particles becoming non-relativistic during matter domination, as it is usually the case for neutrinos, the {\em comoving} free streaming length, $\lambda_{\rm FS}/a$, is actually decreasing in time, and therefore assumes the largest value at the time of the non-relativistic transition. This peculiar distance corresponds to the wave-number
\be
k_{nr} = k_{\rm FS} (z_{nr})\simeq 0.018\,\Omega_m^{1/2}\,\left(\frac{1\,{\rm eV}}{m_\nu}\right)^{1/2}\kMpc\,.
\ee
This scale is typically larger than the scale at which nonlinear effects manifest themselves at low redshifts. At any redshift, scales larger than $1/k_{nr} $ are affected by the presence of massive neutrinos, and, in first approximation, we can write
\be 
\label{eq:Pcc}
 P_{mm}(k) \simeq \begin{cases}
           P_{cc}(k) & \text{for}~ k \ll k_{nr} \\
           (1-f_{\nu})^2\,P_{cc}(k) & \text{for}~ k\gg k_{nr}\,,
          \end{cases}
\ee 
while the exact value of the damping scale will retain a residual redshift dependence. In the left panel of \fig{fig:PL} we plot the ratio $P_{mm}/P_{cc}$ for \smnu$=0.3,\,0.53$ eV at redshifts $z=0,\,2$, showing the two asymptotic regimes of \eq{eq:Pcc}. On very large scales the ratio goes to one, while at small scales it approaches $(1-f_{\nu})^{2}$ regardless of the redshift. Intermediate scales are instead affected by the actual value of the free streaming wave number $k_{FS}(m_{\nu},z)\equiv 2\pi a/\lambda_{\rm FS}(m_\nu,z)$, and by its redshift dependence.

It can be shown \cite{HuEisensteinTegmark1998} that, with respect to the massless case, the total {\em linear} power spectrum, $P_{mm}$, in massive neutrino scenarios is asymptotically suppressed at $z=0$ by a constant factor on scales $k \gg k_{nr}$
\be
 \frac{P_{mm}(k;f_\nu)}{P_{mm}(k;f_\nu = 0)} \simeq 1-8 f_{\nu}\,,
\ee
while from \eq{eq:Pcc} and \eq{eq:Pmm}, it follows that the suppression for the CDM power spectrum, $P_{cc}$, is given by a factor $\sim (1-6 f_{\nu})$. The difference in the suppression between the two power spectra is shown in the right panel of \fig{fig:PL}.

\section{Simulations}
\label{sec:sim}

The DEMNUni simulations have been conceived for the testing of different probes, including galaxy surveys, CMB lensing, and their cross-correlations, in the presence of massive neutrinos. To this aim, this set of simulations is characterised by a volume big enough to include the very large-scale perturbation modes, and, at the same time, by a good mass resolution to investigate small-scales nonlinearity and neutrino free streaming.  Moreover, for the accurate reconstruction of the light-cone back to the starting redshift of the simulations, it has been used an output-time spacing small enough that possible systematic errors, due to the interpolation between neighbouring redshifts along the line of sight, result to be negligible.

The simulations have been performed using the tree particle mesh-smoothed particle hydrodynamics (TreePM-SPH) code \gadgetthree, an improved version of the code described in \cite{Springel2005}, specifically modified in \cite{VielHaehneltSpringel2010} to account for the presence of massive neutrinos. This version of \gadgetthree\ follows the evolution of CDM and neutrino particles, treating them as two distinct collisionless fluids. For the specific case of the DEMNUni simulations, a \gadgetthree\ version,  modified for OpenMP parallelism and for memory efficiency, has been used to smoothly run on the BG/Q Fermi cluster. 

Given the relatively high velocity dispersion, neutrinos have a characteristic clustering scale larger than the CDM one. This allows to save computational time by neglecting the calculation of the short-range tree-force  induced by the neutrino component. This results in a different scale resolution for the two components, which for neutrinos is fixed by the PM grid (chosen with a number of cells eight times larger than the number of particles), while for CDM particles is larger and given by the tree-force (for more details see \cite{VielHaehneltSpringel2010} ).  This choice does not affect the scales we are interested in, since, as shown in \cite{VillaescusaEtal2013}, the application of the short-range tree force to neutrino particles is required only to describe the neutrino density profile inside massive halos at low redshifts. In this work we are not considering this kind of analysis on such small scales. 

Each DEMNUni run starts at redshift $z_{in}=99$, and is characterised by a comoving volume of $8\cGpc$, filled with 2048\cub\ dark matter particles and, when present, 2048\cub\ neutrino particles. Given the large amount of memory required by the simulations, baryon physics is not included. Nevertheless, since we are looking for deviations from a fiducial reference $\Lambda$CDM model, it is expected that baryon feedback cancels out when considering relative effects, \emph{i.e.} the massive with respect to the massless neutrino case. In addition we expect also that any additional effect produced  by the interplay of neutrinos with baryon physics should be of higher order. This is supported also from \cite{BirdVielHaehnelt2012} which shows that the neutrino induced suppression in the total matter power spectrum is very much the same also when neutrinos are considered in the presence of baryons. Therefore, our choice should not affect the results presented in this work.

\begin{table}[t]
\begin{center}
\begin{tabular}[]{l | c c  c c c  c c}
&  $\sum m_{\nu}$[eV]   & $\Om_{cdm}$ & $f_{\nu}$ & $\sigem$ & $\sigec$ &$m^{c}_p [\Ms]$ &$m^{\nu}_p [\Ms]$\\                    
\noalign{\smallskip}\hline \hline \noalign{\smallskip}
S1       & 0.00 & $0.2700$ & $0.000$ & $0.846$ &$0.846$ & $8.27 \times 10^{10}$ & $-$\\
S2       & 0.17 & $0.2659$ & $0.013$ & $0.803$ &$0.813$ & $8.16 \times 10^{10}$ & $1.05 \times 10^{9}$\\
S3       & 0.30 & $0.2628$ & $0.022$ & $0.770$ &$0.786$ & $8.08 \times 10^{10}$ & $1.85 \times 10^{9}$\\
S4       & 0.53 & $0.2573$ & $0.040$ & $0.717$ &$0.740$ & $7.94 \times 10^{10}$ & $2.28 \times 10^{9}$\\
\noalign{\smallskip}\hline \hline \noalign{\smallskip}
\end{tabular}
\caption{\label{tab:par} Summary of cosmological parameters and derived quantities for the four models assumed for the DEMNUni simulations. The values $\Omega_b = 0.05$, $\Omega_{m}=0.32$, $h=0.67$, $n_s =0.96$ are shared by all the models.}
\end{center}
\end{table}

We have produced a total of four different DEMNUni simulations, choosing the cosmological parameters according to the \planck\ 2013 results  \cite{Planck2013parameters}, namely a flat \lcdm\ model generalised to a \lcdmn\ , \ie\ a massive neutrino model, by varying only the sum of the neutrino masses over the values \smnut\ (and consequently the corresponding values of $\Omega_\nu$ and $\Omega_c$, while keeping fixed \omegam\ and the amplitude of primordial curvature perturbations $A_{\rm s}$). Table \ref{tab:par} provides a summary of the cosmological parameters that characterise the different runs. It shows as well the derived quantities  $\sigem$ and $\sigec$ corresponding to the r.m.s. of perturbations on spheres of $8\Mpc$, computed respectively for the total and CDM matter components, and the CDM and neutrino mass particle resolutions, which vary according to the value of $\Omega_c$ and $\Omega_\nu$.

The simulations are characterised by a softening length $\varepsilon=20$\hmone\ kpc, and have been run on the Fermi IBM BG/Q supercomputer at CINECA\footnote{Consorzio Interuniversitario del Nord-Est per il Calcolo Automatico, Bologna, Italy: \href{http://www.hpc.cineca.it/services/iscra}{http://www.hpc.cineca.it/services/iscra}.}, employing about 1 Million cpu-hrs per simulation (including the production of halo and sub-halo catalogues). For each simulation, 62 outputs have been produced, logarithmically equispaced in the scale factor $a=1/(1+z)$, in the redshift interval $z=0-99$, 49 of which lay between $z=0$ and $z=10$. For each of the 62 output times, it has been produced on-the-fly one particle snapshot, composed by both CDM and neutrino particles, one 3D Cartesian grid of the gravitational potential, $\phi$, and one 3D cartesian grid of its time derivative, $\dot{\phi}$, with a mesh of dimension 4096\cub\ that covers a comoving volume of 2 \hmone\ Gpc, for a total of about 90 TB of data per simulation. In addition, in order to build parent-halo catalogues, the simulation outputs have been first processed with the friends-of-friends (FoF) algorithm included in the \gadgetthree\ package. The code is applied only to CDM particles with linking length set to 0.2 times the mean inter-particle distance.
A minimum number of 32 particles to identify a structure has been assumed, fixing the minimum parent-halo mass to $M_{\rm FoF}\simeq2.5\times 10^{12}\Ms$. 
The FoF catalogues have subsequently been processed with the \subfind\ algorithm \cite{SpringelYoshidaWhite2001, DolagEtal2009}, which identifies locally overdense, gravitationally bound regions within an input parent halo. With this procedure some of the initial FoF parent halos are split into multiple sub-halos, with the result of an increase of the total number of identified objects, and of a lower minimum mass limit. Here, a minimum number of $20$ particles has been adopted in order to constitute a valid sub-halo. Moreover, by means of a specific routine included in the \subfind\ algorithm, spherical overdensity halo catalogues have been produced, and, in particular, for each halo the value of \rfiveh\ has been computed; this is defined as the radius enclosing a matter (CDM+neutrinos) density equal to 200 times the mean density of the Universe $\rho_{m}(z)$ at redshift $z$, and the corresponding mass in terms of $M_{\rm 200}\equiv 200\,\rho_m\,\frac{4\,\pi}{3}\,R_{\rm 200}^{\,3}$. We stress here that all the post-processing has been modified to take the neutrino component properly into account.
\begin{figure}[t!]
\begin{center}
\includegraphics[width=.55\textwidth]{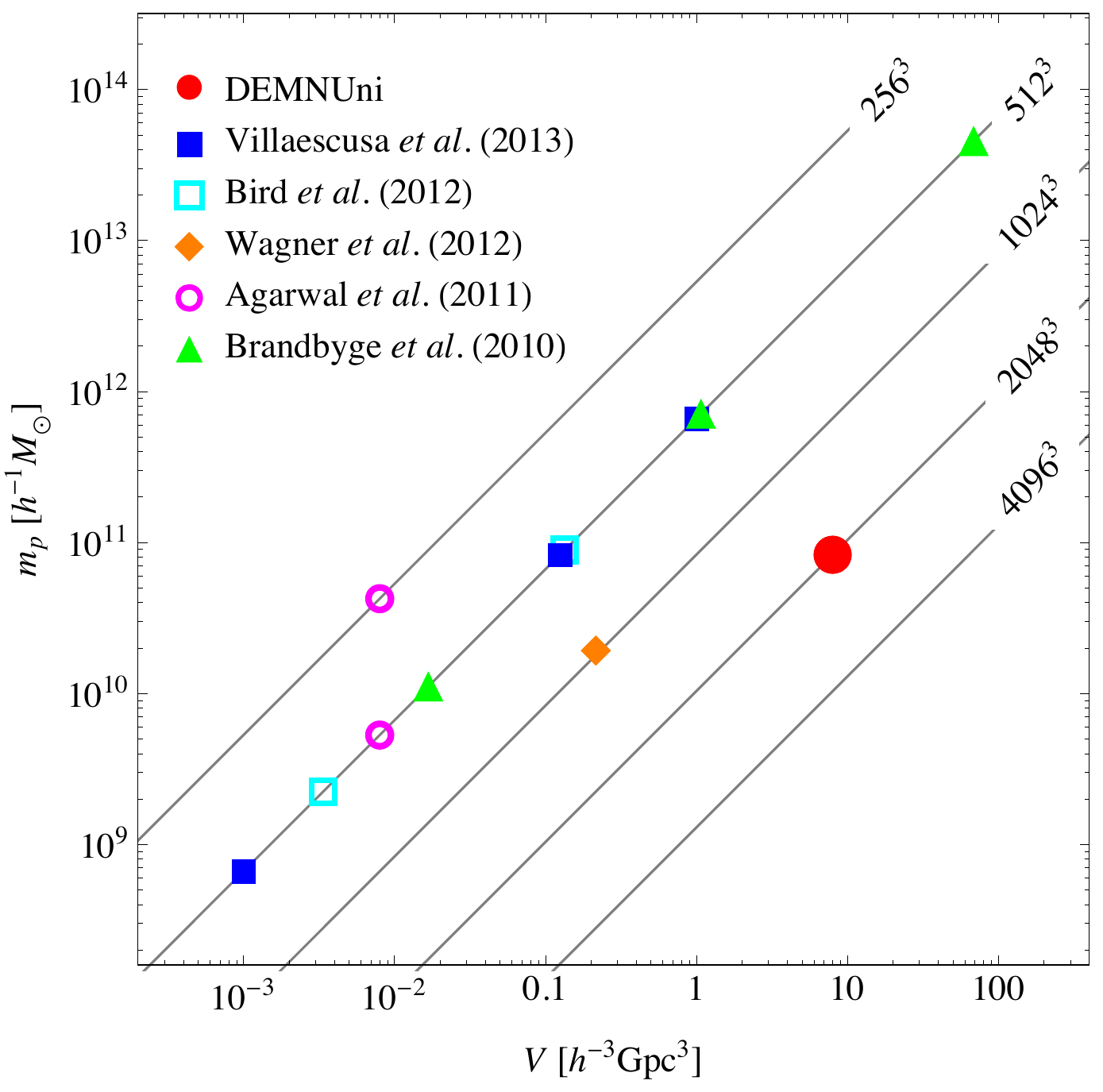}
\caption{\label{fig:sims} 
Comparison between the DEMNUni runs and previous, recent simulations of massive neutrino cosmologies in terms of CDM particle mass resolution and simulation volume. Grey diagonal lines indicate the number of CDM particles.}
\end{center}
\end{figure}

Figure~\ref{fig:sims} presents a comparison of the DEMNUni simulations to previous N-body simulations with massive neutrino particles, in terms of the CDM particle mass resolution $m_p$ and simulation volume $V$. Diagonal grey lines indicate the number of CDM particles. With respect to previous simulations, the DEMNUni suite represents an improvement of about an order of magnitude in terms of particle number (only \cite{BrandbygeEtal2010} considered a larger box, but with considerably smaller mass  resolution).

\section{Matter power spectra}
\label{sec:matterps}

As shown in \S\ref{sec:linear_nu}, the shape of the linear power spectrum is quite sensitive to the value of neutrino masses, and this dependence becomes even stronger in the mildly and fully nonlinear regimes~\cite{BrandbygeEtal2010, BirdVielHaehnelt2012}. Taking advantage of the large DEMNUni simulations volume, in this section we aim at testing the accuracy of current analytical predictions for the nonlinear matter power spectrum, $P_{mm}$, in the presence of massive neutrinos. To this end, we measure individually the different components to $P_{mm}$ in \eq{eq:Pmm}, from very large scales, $k \sim 0.003 \kMpc$, down to fully nonlinear scales, $k\sim 3 \kMpc$, and compare these measurements with PT predictions, in the mildly nonlinear regime, and fitting functions as {\halofit}~\cite{SmithEtal2003, TakahashiEtal2012}, in the fully nonlinear regime\footnote{While the mass resolution of the DEMNUni simulations would allow to look at much smaller scales, of the order of $k\sim 10\kMpc$, we do not investigate this regime since it is dominated by baryon physics~\cite{RuddZentnerKravtsov2008, GuilletTeyssierColombi2010, CasariniEtal2011B, VanDaalenEtal2011, VanDaalenEtal2013}.}. The goal here is to understand if possible departures from the linear regime of neutrino perturbations have to be taken into account for precision cosmology at the \% level. 

\begin{figure}[t]
\begin{center}
\includegraphics[width=.98\textwidth]{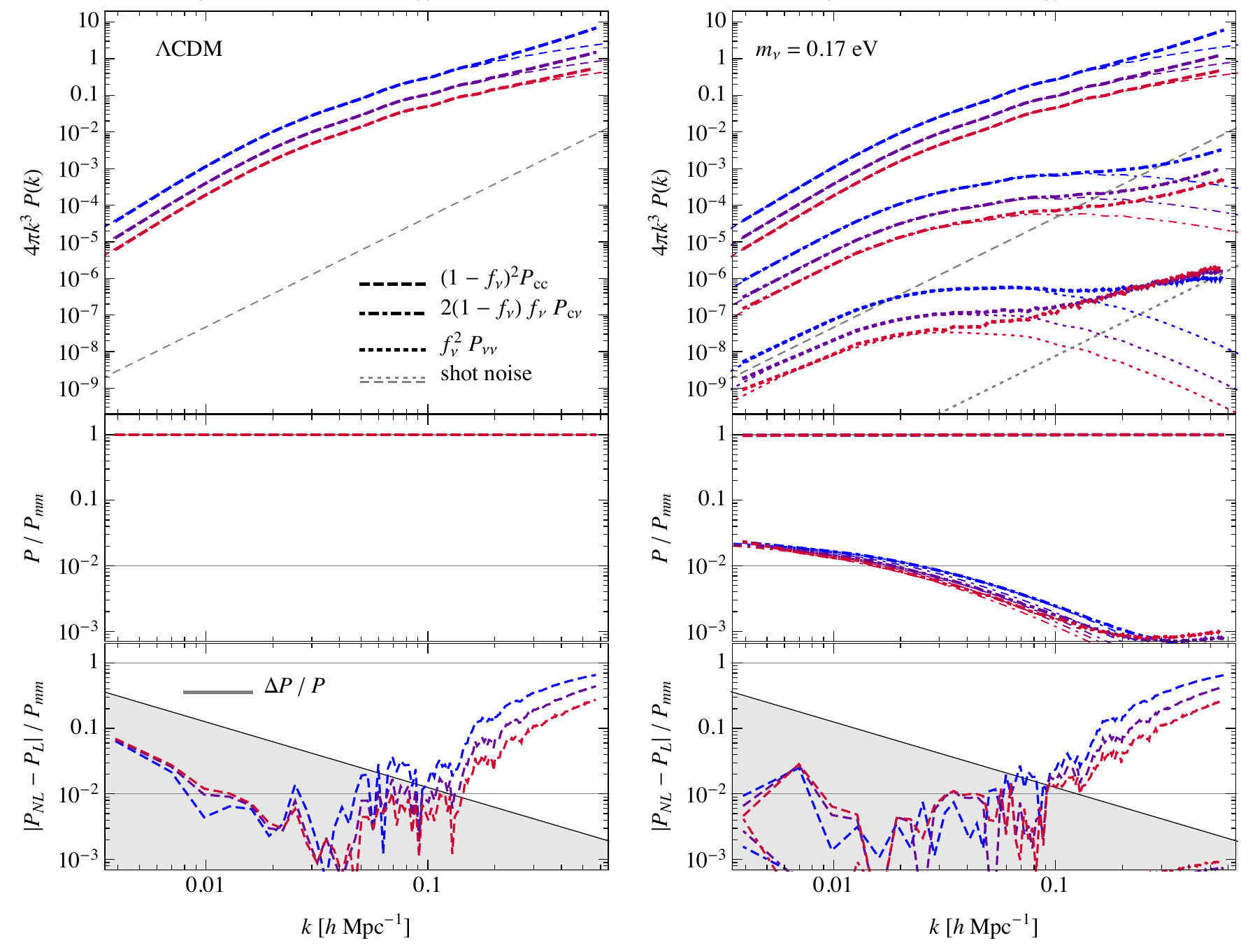}
\end{center} 
\caption{\label{fig:psnlA} Comparison between the different contributions to the nonlinear matter power spectrum, $(1-f_\nu)^2\,P_{cc}$ ({\em dashed curves}), $f_\nu^2\,P_{\nu\nu}$ ({\em dotted}) and $2\,f_\nu\,(1-f_\nu)\,P_{c\nu}$ ({\em dot-dashed}), as described in the text. All the measurements at redshifts $z=0,\,1,\,2$ are shown with shades varying, respectively, from blue to red. Thin coloured curves correspond to the respective linear predictions. Dashed and dotted grey lines on the top panels show the shot-noise contributions to the CDM and neutrinos power spectra. The shaded area in the bottom panel shows values below the 1-$\sigma$, relative, Gaussian uncertainty on $\Delta P(k)/P(k)=1/\sqrt{2\pi k^2/k_f^2}$, $k_f$ being the fundamental frequency of the simulation box.}   
\end{figure}
\begin{figure}[t]
\begin{center}
\includegraphics[width=.98\textwidth]{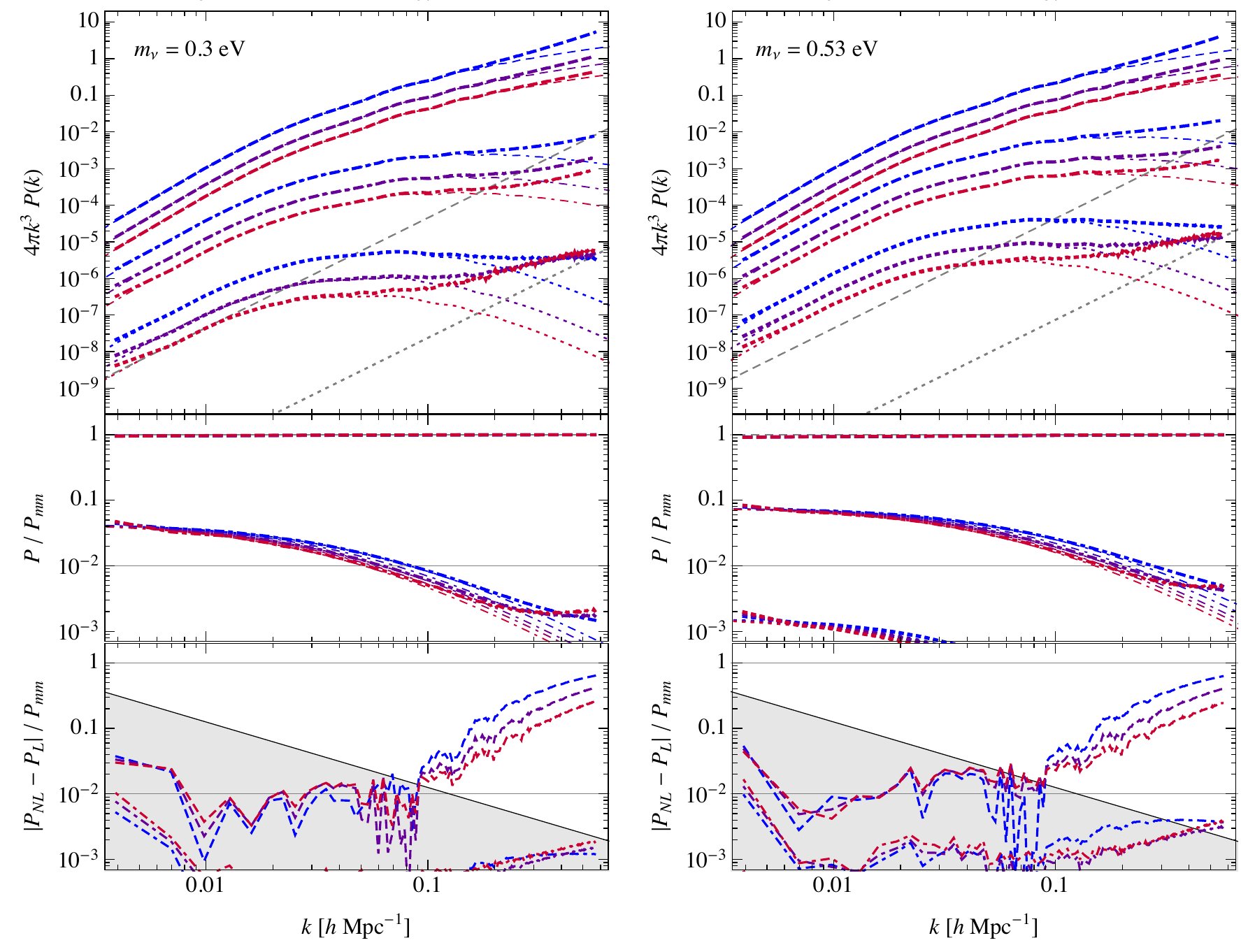}
\end{center}
\caption{\label{fig:psnlB} Same as figure~\ref{fig:psnlA}, but for \smnu$=0.3,\,0.53$ eV.}   
\end{figure}
Before proceeding to the comparison with the nonlinear, analytical predictions, however, we take a look at each component and its relative contribution to the nonlinear $P_{mm}$ measured from the simulations. \figs{fig:psnlA} and \ref{fig:psnlB} show the CDM auto power spectrum, $(1-f_\nu)^2\,P_{cc}$ ({\em dashed curves}), the neutrino auto power spectrum, $f_\nu^2\,P_{\nu\nu}$ ({\em dotted curves}), and the cross CDM-neutrino power spectrum, $2\,f_{\nu}\,(1-f_\nu)\,P_{c\nu}$ ({\em dot-dashed}), as extracted from the simulations ({\em thick curves}), and the corresponding linear predictions ({\em thin curves}). Each plot corresponds to a different value of the neutrino mass in the simulations. Within each plot, the top panel shows the adimensional power spectrum, $4\,\pi\,k^3\,P(k)$, the middle panel the ratio of each contribution to $P_{mm}$, in the nonlinear and linear cases, and the bottom panel shows the ratio $(P_{NL}-P_L)/P_{mm}$, \ie the difference between each measured nonlinear component, $P_{NL}$, in \eq{eq:Pmm}, and the corresponding linear prediction, $P_{L}$, compared to the nonlinear $P_{mm}$. All the measurements at redshifts $z=0,\,1,\,2$ are shown with different shades varying from blue to red. Dashed and dotted grey lines on the top panels show the shot-noise contributions to the CDM and neutrinos power spectra, respectively. The shaded area in the bottom panel show the region below the 1-$\sigma$, Gaussian uncertainty on $P(k)$, given by $\Delta^2 P(k)=P^2(k)/(2\pi k^2/k_f^2)$, $k_f$ being the fundamental frequency of the simulation box (the shot-noise contribution to the variance is ignored for simplicity).

In the linear regime, $k\lesssim 0.1 \kMpc$, most of the power comes from the CDM component, with the cross power spectrum term accounting roughly for a 10\% of the total $P_{mm}$, and the neutrinos being already negligible (see middle panels of \figs{fig:psnlA} and \ref{fig:psnlB}). On nonlinear scales, from the lower panels we notice that most of the nonlinear contribution to $P_{mm}$ is still given by $P_{cc}$, which is the only component significantly deviating from the linear prediction, while the other two terms remain linear, and, therefore, less and less important as we move to smaller scales\footnote{Let us notice that measurements of the neutrino auto power spectrum and neutrino-cold matter cross power spectrum shown in both \figs{fig:psnlA} and \ref{fig:psnlB}, present a spurious contribution at small scales {\em not} to be confused with a residual shot noise component. Such contribution, which scales roughly (but not exactly) as $1/k$, is due to the neutrino velocity distribution and it does not grow with time, so that at low redshift is overtaken by the expected neutrino power spectrum. A detailed discussion of numerical issues associated with the set-up of the initial conditions will be discussed in a forthcoming work \cite{BelEtal2015}.}. It should be kept in mind that each contribution in \figs{fig:psnlA} and \ref{fig:psnlB} is multiplied by the proper power of $f_\nu$ or $(1-f_\nu)$, with $f_{\nu}\simeq \mathcal{O} (2-4 \%)$ .
We therefore conclude that, on the scales probed by the DEMNUni simulations, and considering the present constraints on the sum of neutrino masses, {\em the total nonlinear matter power spectrum, in massive neutrino cosmologies, can be described at the 1\% level by accounting for the nonlinear evolution of CDM perturbations alone, while adopting the linear prediction for the other components}. This result will be useful for analytical predictions of the nonlinear $P_{mm}$ discussed in the next sections.

\subsection{Perturbation Theory}
\label{ssec:pt}

For the first time in the literature, the large volume of the DEMNUni simulations gives us the possibility to measure the matter power spectrum in neutrino cosmologies at the 1\% accuracy level, on a very large range of scales, allowing a test of PT predictions at the accuracy level required by current and futures galaxy surveys. 

Several works in the literature have discussed the effects of massive neutrinos in cosmological perturbation theory beyond the linear level \cite{SaitoTakadaTaruya2008, Wong2008, SaitoTakadaTaruya2009, LesgourguesEtal2009, ShojiKomatsu2009, ShojiKomatsu2010, SaitoTakadaTaruya2011, FuhrerWong2015,BlasEtal2014}. In these descriptions, the neutrino component is treated, similarly to the CDM one, as a single perfect fluid, fully characterised in terms of its density and velocity divergence (see, however, \cite{DupuyBernardeau2014, DupuyBernardeau2015} for a multiple-flow approach to the evolution of neutrino perturbations). The  main difference with respect to the $\Lambda$CDM case is represented by an effective sound speed modifying the Euler equation for the neutrino component, and accounting for the neutrino velocity distribution. The first consequence, at the linear level, is a scale-dependent linear growth factor, $D(k,z)$, for both the CDM and neutrino components. However, the perfect-fluid approximation fails to provide an accuracy for the neutrino power spectrum below the 10\% level \cite{ShojiKomatsu2010}. Nonetheless, as shown in \S \ref{sec:matterps}, since the neutrino contribution to the total matter power spectrum is order of magnitudes smaller than the CDM one, such discrepancies on the neutrino component alone do not affect significantly the CDM and total matter power spectra. Therefore, we will assume the two-fluid approximation for all the comparisons of analytical versus numerical results in this section.

In addition, even if in the mildly nonlinear regime the effective sound speed affects as well the mode-coupling at all the orders of the perturbative expansion, we will follow the same approach adopted by \cite{SaitoTakadaTaruya2009}. They have shown that limiting the neutrino-induced scale-dependence to the linear growth factor alone (and, therefore, the use of standard EdS- like kernels in the perturbative expansion) proves to be a quite good approximation to the full PT solution for the nonlinear CDM field, on scales where PT is expected to be accurate. 

Finally, we will make the additional approximation, already proposed in \cite{SaitoTakadaTaruya2008}, of describing the neutrino perturbations by means of their linear solution. While this is not {\em per se} a good assumption \cite{BlasEtal2014, FuhrerWong2015}, it does provide the correct neutrino contribution to the total matter power spectrum on the (large) scales where such contribution is relevant. 

As a starting point for future, more accurate comparisons, we will consider, therefore, the following perturbative prediction for the total matter power spectrum
\be
\label{eq:PT_mm}
P_{mm}^{PT}(k)=(1-f_\nu)^2\,P_{cc}^{PT}(k)+2\,(1-f_\nu)\,f_{\nu}\,P_{c\nu}^L(k)+f_\nu^2\,P_{\nu\nu}^L(k)\,.
\ee
Here, the contribution $P_{cc}^{PT}(k,z)$ represents the nonlinear power spectrum predicted in perturbation theory along the lines of \cite{SaitoTakadaTaruya2009}, \ie it is computed in terms of its linear counterpart, $P_{cc}^{L}(k,z)$, which provides the correct {\em linear} scale-dependence of the growth factor, but assumes the standard EdS nonlinear kernels in the perturbative expansion. Differently from previous works, however, we do not only consider standard, one-loop corrections to $P_{cc}^{L}$, but we take into account also standard PT two-loop corrections, as well as the ``regularised'' predictions based on the multi-point propagator expansion of \cite{BernardeauCrocceScoccimarro2008}, computed using the \textsc{RegPT} code of \cite{TaruyaEtal2012}. Since the \textsc{RegPT} code does not account for the evolution of the scale-dependent linear growth in massive neutrino models, we produce all the predictions at $z>0$ by providing the corresponding $z$-input linear power spectrum as a ``fake $z=0$ input'' required by the code. 
\begin{figure}[t]
\includegraphics[width=1\textwidth,center]{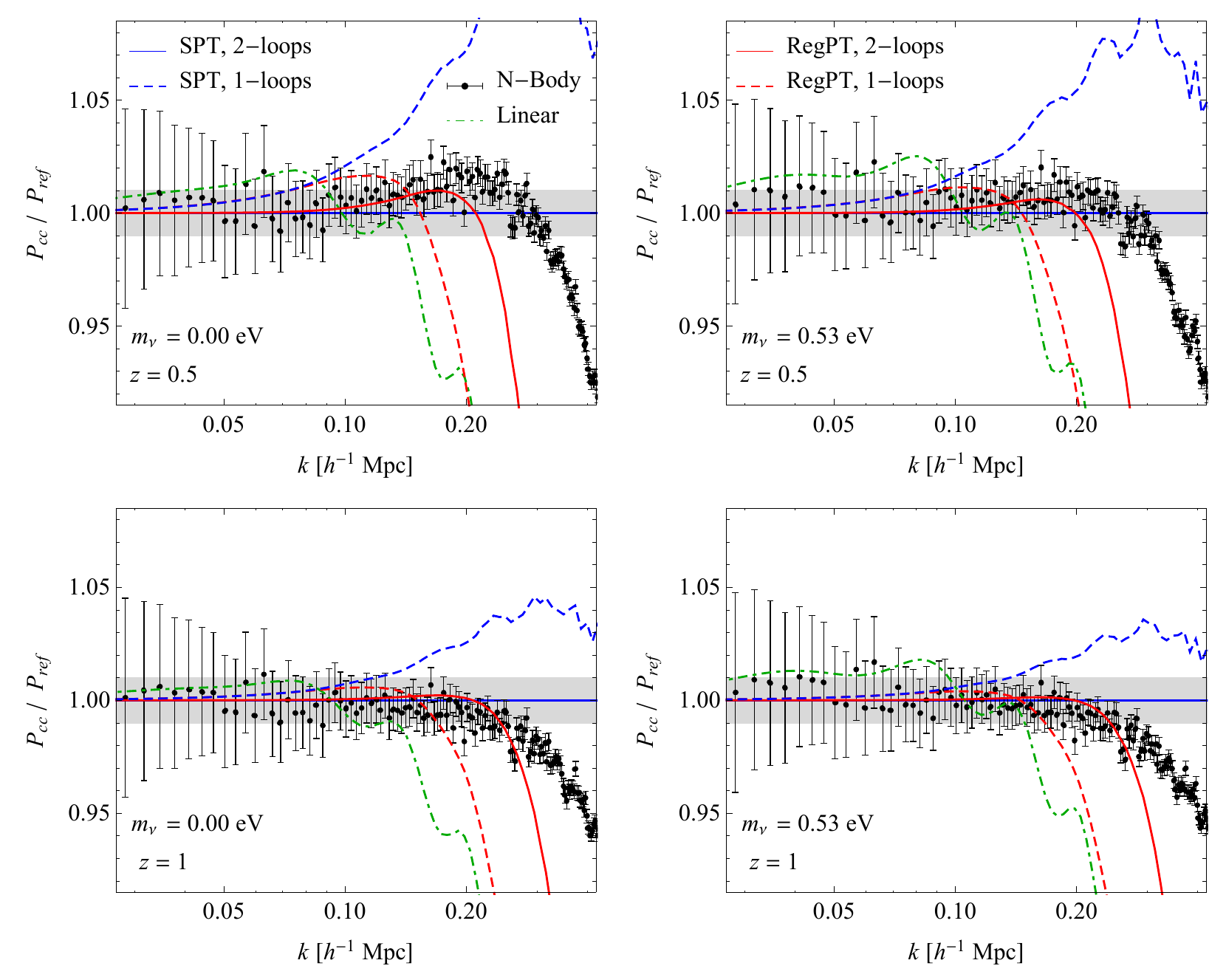}
\caption{\label{fig:ps_PT_CDM} 
Perturbation Theory predictions for the cold matter power spectrum $P_{cc}(k)$. Each panel shows the measurements from the N-body simulations, divided by the reference power spectrum given by the two-loop, standard PT results ({\em black points with error-bars}). Also shown are the corresponding ratios for the linear ({\em green, dotted}), one-loop, standard PT ({\em blue, thin, dashed curve}), multi-point propagator expansion at one- ({\em red, thick, dashed}) and two-loops ({\em red, thick, continuous}) ass obtained from the \textsc{RegPT} code of \cite{TaruyaEtal2012}.}
\end{figure}

While this is not the most rigorous approach, it represents a practical application, to massive neutrino scenarios, of available tools developed within the $\Lambda$CDM framework. As we will see, the gain in accuracy achieved by recent resummation schemes, applied here to the CDM component alone, might compensate for the crude approximations that this approach implies. Clearly we are only considering predictions for the CDM and total matter power spectra, as these statistics are the relevant ones for galaxy clustering and weak lensing observations.

In \fig{fig:ps_PT_CDM} we show the perturbative results against the measurements at $z=0.5,\,1$, and for \smnu$=0,\,0.53$ eV. Error bars are the theoretical expectation for a Gaussian field, that is
\be
\Delta P^2(k)=\frac{1}{2\pi\,k^2\,k_f}\,\left[P(k)+\frac1{(2\,\pi)^3\,\bar{n}}\right]^2\,,
\ee
where $k_f\equiv 2\pi/L$ is the fundamental frequency of the simulation box, $L$ being its linear size, and $\bar{n}$ is the particle number density accounting for the shot-noise component\footnote{The relatively small scatter of data points with respect to the error bars is due to the specific seed chosen for the random number generator used for the set-up of the initial conditions \cite{2015arXiv150305920S}.}. 

Let us first notice that, in the $\Lambda$CDM case (left panels in \fig{fig:ps_PT_CDM}), the two-loops standard PT does not provide a good fit to the data at low redshifts \cite{BlasGarnyKonstandin2014}, while it reproduces fairly well the simulation measurements at $z\ge1$. Analytic predictions are 1\%  accurate at $z=1$, up to a maximum wave-number $k_{max} \simeq 0.3\kMpc\;$\footnote{It is worth noticing that the agreement may also depend on the simulation mass resolution; we expect that much higher resolutions lead to more power at small scales \cite{FosalbaEtal2013A}.}. 
\begin{figure}[t]
\begin{center}
\includegraphics[width=1.\textwidth,center]{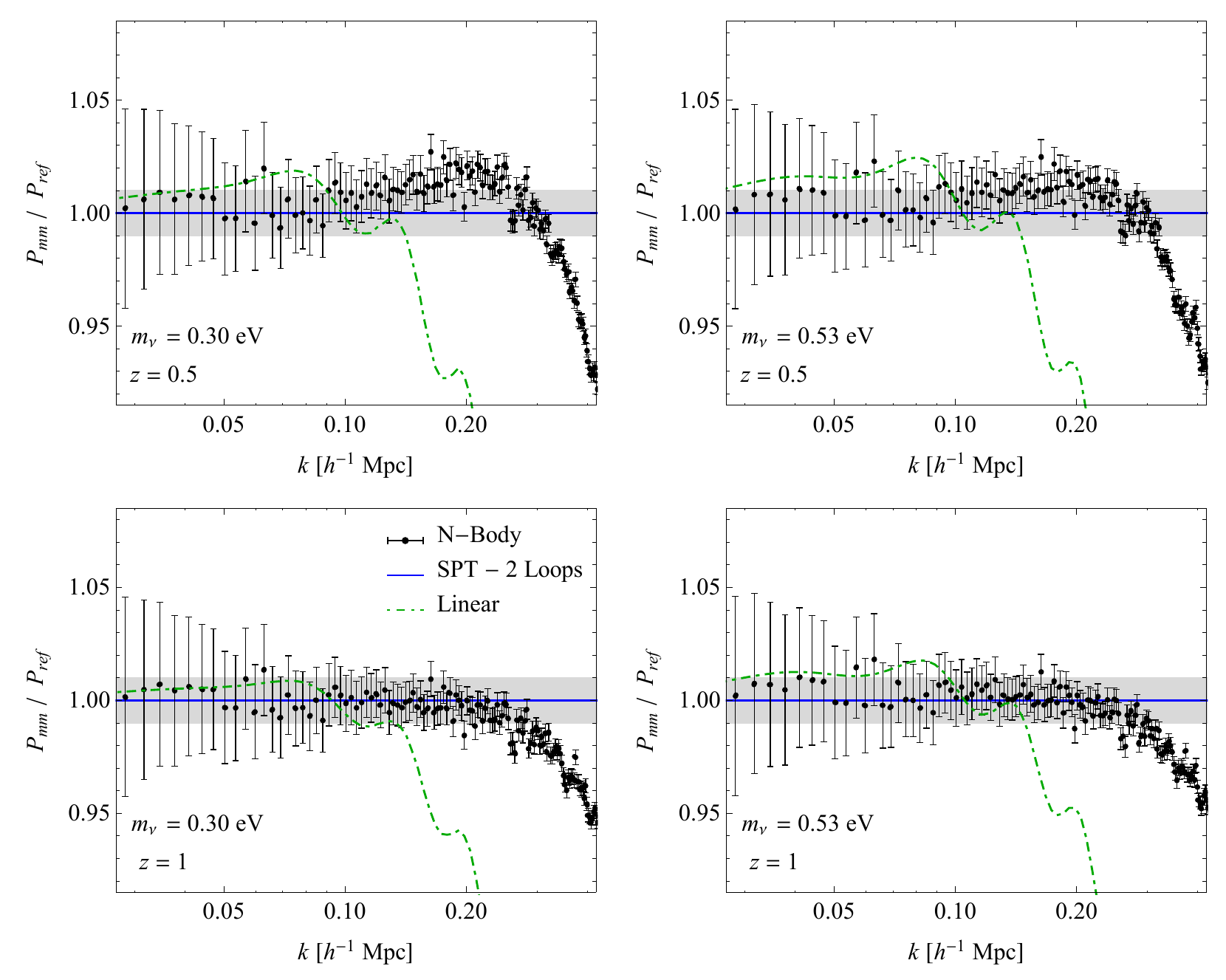}
\caption{\label{fig:ps_PT_DM} 
Same as \fig{fig:ps_PT_CDM} but for the total matter power spectrum with \smnu$=0.3$ eV ({\em left panels}) and 0.53 eV ({\em right panels}). Theoretical predictions have been obtained using \eq{eq:PT_mm}, \ie computing nonlinear correction for CDM only.}
\end{center}
\end{figure}

Turning our attention to the CDM power spectrum in massive neutrino cosmologies (right panels in \fig{fig:ps_PT_CDM}), at $z\ge1$ we find approximately that the $1\%$ accuracy is recovered up to a value of $k_{max}$ very close to the $\Lambda$CDM case.
On the other hand, at {\em lower redshifts}, $z<1$, we notice that, with respect to the massless case, PT predictions are in better agreement with the measurements extracted from the DEMNUni simulations. This is a consequence of the fact that, due to massive neutrino free streaming, the nonlinear evolution of CDM perturbations is suppressed, and therefore the range of scales in which PT corrections show the same accuracy as in standard $\Lambda$CDM cosmologies increases with $f_\nu$, \ie with the relative contribution of the total neutrino mass to $\Omega_m$. In addition, since, massive neutrino free streaming has an impact already at the linear level, these models are characterised by a lower value of $\sigma_{8}$ (and $\sigma_{8,cc}$) than a $\Lambda$CDM universe with the same $\Omega_m$.

Given a prediction for the CDM power spectrum accurate at the 1\% level up to a given $k_{max}$, we check if the perturpative approach of \eq{eq:PT_mm} reproduces, with the same level of accuracy, the total matter power spectrum measured from the simulations. The results are illustrated in \fig{fig:ps_PT_DM}, which shows that, indeed, the linear treatment of the $P_{c \nu}$ and $P_{\nu}$ contributions to the total $P_{mm}$ proves to be a very good approximation. The difference in the accuracy of the predictions between the \smnu$=0.3$ eV ({\em left panels}) and the \smnu$=0.53$ eV ({\em right panels}) cases is again mainly due to the different values of $f_\nu$ and, therefore, to the different effect of neutrino free streaming on dark matter perturbations, according to the total neutrino mass: for a given value of $\Omega_m$, a larger value of \smnu\ not only increases the relative amount of neutrino perturbations that are washed out below the free streaming scale, $\lambda_{\rm FS}$, consequently reducing the contribution of $P_{\nu}$ and $P_{c \nu}$ to the total $P_{mm}$, but also decreases the factor $(1-f_\nu)^2$ in front of $P_{cc}$ in \eq{eq:PT_mm}, where the nonlinear evolution of $P_{cc}$ is in turn suppressed, with respect to the massless case, by the action of the total gravitational potential sourced both by CDM and massive neutrinos. Apart from inducing a scale-dependence of the linear growth factors for CDM and total matter, the main direct product of this physical mechanism is represented by a lower amplitude of linear perturbations at $z=0$, where $\sigma_{8,cc}= 0.786,\,0.740$, and $\sigma_{8,mm}=0.770,\,0.717$, for \smnu$=0.3,\,0.53$ eV, respectively. We will show in \S\ref{ssec:halofit} that, on scales much smaller than the so-called turn-over scale, beyond the mildly nonlinear regime, $k>0.2\,\kMpc$, where the growth factor scale-dependence induced by neutrino free streaming approaches its asymptotic value depending only on $f_\nu$ (see \eq{eq:fc_small} in \S\ref{ssec:rsd}), the effect of massive neutrinos on $P_{cc}$ and $P_{mm}$ mostly reduces to a mere rescaling of the power spectra in the massless case, according to the values of  $\sigma_{8,cc}$ and $\sigma_{8,mm}$.

Recently the BOSS collaboration released new constraints on neutrino masses based on measurements of the galaxy power spectrum multipoles in redshift space \cite{BeutlerEtal2014} at the mean redshift $z=0.57$. The BOSS analysis was based on a prediction for the total matter power spectrum computed applying standard PT directly to the linear $P_{mm}$. The outcome of such a calculation, in principle even less theoretically justified than our crude assumptions, is very similar to our results obtained via \eq{eq:PT_mm}; we have checked that any difference between the two approaches stays below the 1\% level at the scales relevant for current observations. 

A test somehow similar to the one presented here is shown in \cite{UpadhyeEtal2014}, where the authors compare different PT predictions, including the Time-RG method of \cite{Pietroni2008, LesgourguesEtal2009}, to simulations of CDM particles, modifying only the background evolution and the initial conditions to account for free streaming massive neutrinos. They show that PT predictions are in agreement with the measurements of CDM power spectra, extracted from their CMD simulations, at the \% level, over a similar range of scales as tested in this work. However they assume a scale-independent growth rate to rescale back the late time ($z=0$) CDM power spectrum, $P_{cc}(k)$, to the initial redshift of the simulations. By doing so, the linear dynamics cannot be correctly recovered at any $z$ other than $z=0$. 

Finally, the crucial results of this section rely on the discussion of \S\ref{sec:matterps} and the measurements shown in \figs{fig:psnlA} and \ref{fig:psnlB}, that is the contributions from the two terms, other than $P_{cc}$, entering \eq{eq:PT_mm}, and in particular from the cross power spectrum $P_{c\nu}$, always remain negligible compared to $P_{cc}$ on nonlinear scales, at least for the level of accuracy requested for PT to be useful.

\subsection{Fitting functions}
\label{ssec:halofit}

Given the limitations of the perturbative approaches, it is sometimes convenient and/or sufficient to resort to fitting functions for the nonlinear matter power spectrum. In this section we see how the approximation of linear evolution for neutrino perturbation can be applied as well to the {\halofit} prescription \cite{SmithEtal2003}. 

The {\halofit} formula, originally based on stable clustering considerations \cite{HamiltonEtal1991, PeacockDodds1994}, provides a mapping between the linear power spectrum and the nonlinear one, which in turn depends on few cosmological parameters, \eg, $\Omega_m$, and several free parameters determined by comparisons against measurements from N-body simulations. A new, more accurate, version of the fitting formula has been recently presented by \cite{TakahashiEtal2012}. The revised formula is expected to be accurate at the 5\% level for $k<1\kMpc$ and $z\le 10$, while it degrades to the 10\% level for $k<10\kMpc$ and $z\le 3$.
\begin{figure}[t]
\begin{center}
\includegraphics[width=.98\textwidth]{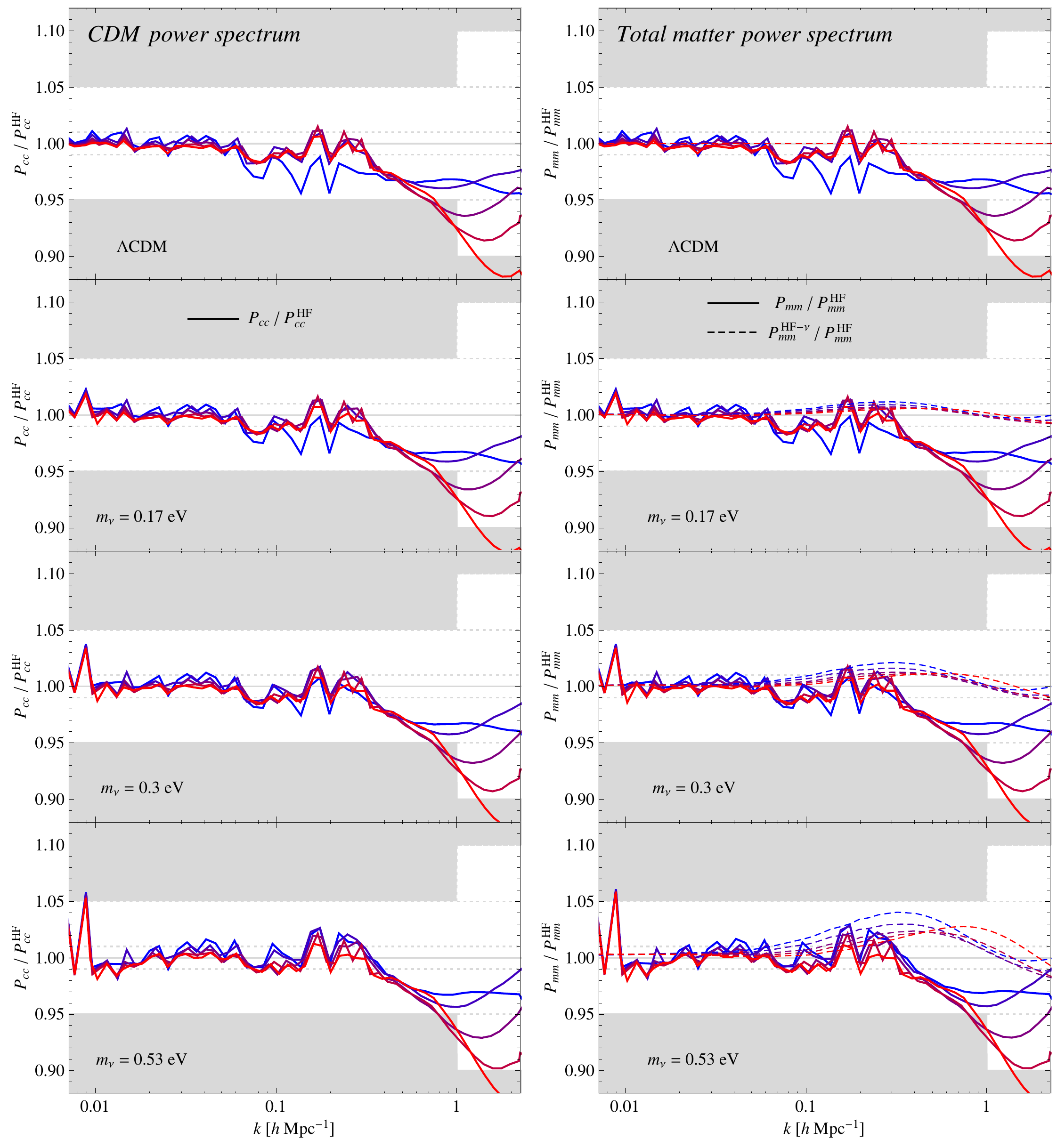}
\caption{\label{fig:ps_HF} 
{\em Left column}: ratio of the measured CDM nonlinear power spectrum, $P_{cc}$, to the {\halofit} prediction $P_{cc,HF}$ from \cite{TakahashiEtal2012}, with no additional parameters to account for neutrino effects. Each panel correspond to one value of \smnu, with different colours denoting different redshifts: $z=0,\,0.5,\,1,\,1.5,\,2$, blue to red. Shaded areas denote the regions beyond the accuracy expected for the formula. {\em Right column}: ratio of the measured total matter nonlinear power spectrum, $P_{mm}$, to the prediction, $P_{mm}^{HF}$, of \eq{eq:halofit}. }
\end{center}
\end{figure}

In the context of massive neutrinos cosmologies, Ref.~\cite{BirdVielHaehnelt2012} provided corrections (controlled by a few, additional parameters) to the original {\halofit} formula, in order to account for neutrinos effects on the nonlinear {\em total} matter power spectrum\footnote{The massive neutrino corrections of \cite{BirdVielHaehnelt2012} have been implemented, along the revised version of \cite{TakahashiEtal2012}, in the most recent versions of the \texttt{camb} \cite{LewisChallinorLasenby2000} and \texttt{class} \cite{Lesgourgues2011} codes.}. However, as shown in \figs{fig:psnlA} and \ref{fig:psnlB}, nonlinear corrections to the cross power spectrum between CDM and neutrinos are below the percent level, therefore we wonder if, similarly to \eq{eq:PT_mm}, a fitting formula for the total matter power spectrum, $P_{mm}(k)$, could be given in terms of the linear predictions, $P_{c\nu}^{L}(k)$ and $P_{\nu\nu}^{L}(k)$, and the {\halofit} fitting formula applied directly to the linear CDM power spectrum, $P_{cc}^{HF}(k)$, that is 
\be\label{eq:halofit}
P_{mm}^{HF}(k)\equiv(1-f_\nu)^2\,P_{cc}^{HF}(k)+2\,f_\nu\,(1-f_\nu)\,P_{c\nu}^L(k)+f_\nu^2\,\,P_{\nu\nu}^L(k)\,.
\ee 
Here $P_{cc}^{HF}(k)\equiv{\mathcal F}_{HF}[P_{cc}^{L}(k)]$, where the {\halofit} mapping, ${\mathcal F}_{HF}$, between linear and nonlinear power spectra does not account for additional corrections due to massive neutrino physics. 

In the left column of \fig{fig:ps_HF} we show the ratio of the measured CDM power spectrum, $P_{cc}(k)$, to the prediction, $P_{cc}^{HF}(k)$. Each row corresponds to a distinct value of \smnu, while each panel shows the value of this ratio at redshifts $z=0,\,0.5,\,1,\,1.5,\,2$, with colour shades ranging from blue to red. The shaded areas denote the regions beyond the accuracy claimed for the revised formula of \cite{TakahashiEtal2012}. The left panels of \fig{fig:ps_HF} show, indeed, not only that $P_{cc}^{HF}(k)$ provides the expected accuracy, but that it works equally well for all the considered values of the total neutrino mass. This is essential to justify our assumption of applying the {\halofit} mapping to the CDM component alone. Here, we stress again that the version of {\halofit} employed for ${\mathcal F}_{HF}[P_{cc}^{L}(k)]$ does not include any effect due to massive neutrinos on the CDM clustering, since here we use the {\halofit} version developed by \cite{TakahashiEtal2012} in the $\Lambda$CDM framework.
This result is similar to that obtained in \S\ref{ssec:pt} for perturbation theory, and it is a crucial step before checking the validity of the assumptions made in \eq{eq:halofit}.

The right column of \fig{fig:ps_HF} presents the ratio of the measured total matter power spectrum, $P_{mm}(k)$, to the prediction $P_{mm}^{HF}(k)$ of \eq{eq:halofit}.  In addition, dashed curves show the inverse ratio of $P_{mm}^{HF}(k)$ to the specific {\halofit} extension to massive neutrino cosmologies of \cite{BirdVielHaehnelt2012}, denoted as $P_{mm}^{HF-\nu}(k)$. We notice that the simple prescription of \eq{eq:halofit}, while avoiding introducing additional parameters to the fitting formula of \cite{TakahashiEtal2012}, remains well within the expected {\halofit} accuracy. On the other hand, the discrepancies between the prediction of \eq{eq:halofit} and the $P_{mm}^{HF-\nu}(k)$ fit of \cite{BirdVielHaehnelt2012} are within 4\%, with the latter {\em ad-hoc} fit not improving particularly over the former.   
\begin{figure}[t]
\includegraphics[width=0.5\textwidth,center]{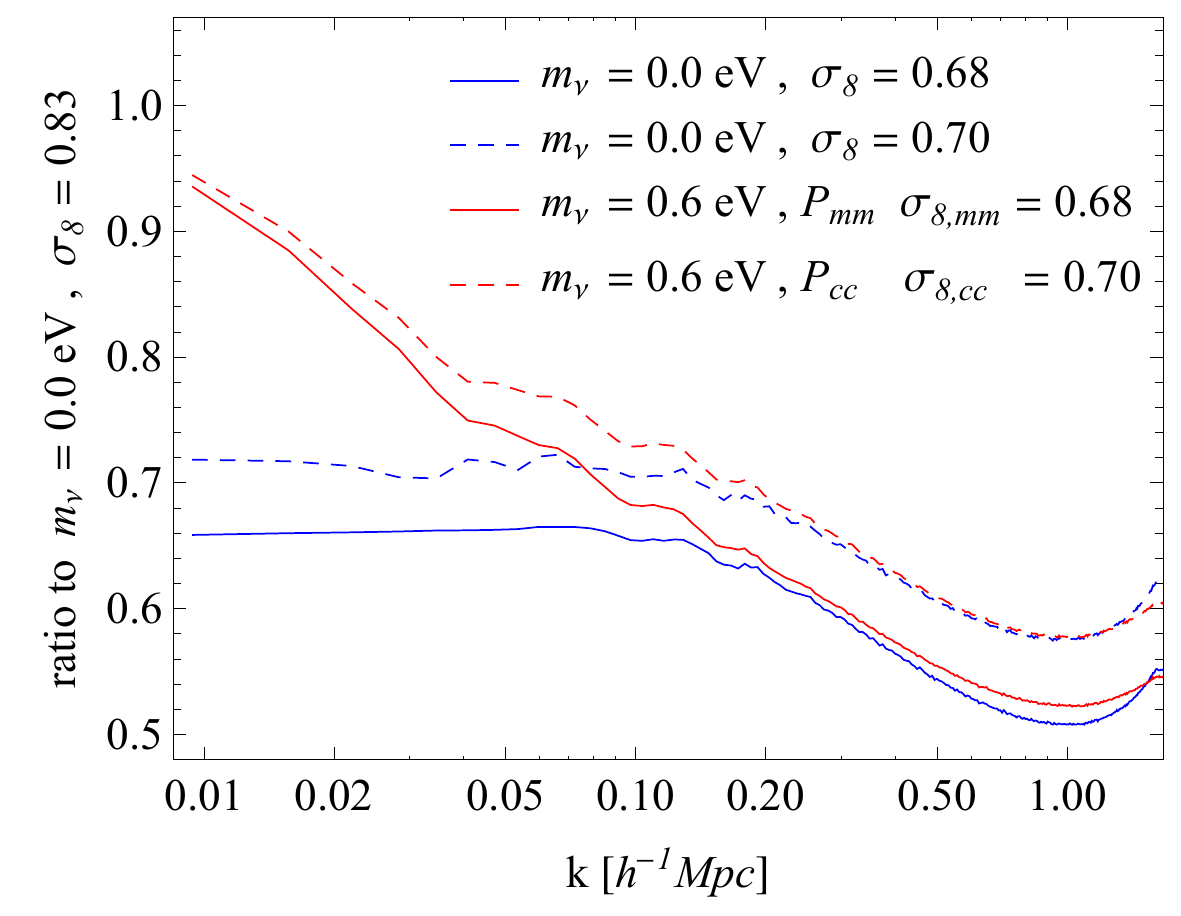}
\caption{\label{fig:cucchiaio} 
Comparison between the nonlinear power spectra measured in two $\Lambda$CDM models ({\em blue curves})  with $\sigma_8=0.68$ ({\em continuous}) and $\sigma_8=0.70$ ({\em dashed}) against the same quantity measured in massive neutrinos models ({\em red curves}) with $\sigma_{8,mm}=0.68$ ({\em continuous}) and $\sigma_{8,cc}=0.70$ ({\em dashed}). All the quantities are shown as ratios to the power spectrum for a $\Lambda$CDM model with $\sigma_8=0.83$ to compare the {\em shape} of the suppression resulting from either a lower overall normalisation or neutrino effects. Measurements for this figure only are from the simulations of \cite{VillaescusaEtal2014}.}
\end{figure}

Here we would like to make some final considerations. The results of this section rely on the fact that, in the first place, in the absence of strong baryon physics, on scales much beyond the mildly nonlinear regime, $k>0.2\,\kMpc$, where the linear growth factor scale-dependence induced by neutrino free streaming approaches its asymptotic value, the extent of the nonlinear evolution of the power spectrum can be accounted for via the amplitude of the linear power spectrum, {\em regardless} of the physical mechanism responsible for the amplitude itself. In other terms, on nonlinear scales, we expect a similar behaviour for the nonlinear matter power spectrum evolved from a linear power spectrum suppressed by massive neutrino free streaming, as from a different linear power spectrum with a lower, primordial normalisation. This assertion can be easily tested with N-body simulations and the results are shown in \fig{fig:cucchiaio}, which makes use, however, of measurements from the simulations described in \cite{VillaescusaEtal2014,Massara_etal_2014}. The plot shows ratios, with respect to the matter power spectrum in a given $\Lambda$CDM cosmology with $\sigma_8=0.83$, of the matter power spectrum in other $\Lambda$CDM cosmologies (with different $\sigma_8$ normalisations ), and of the CDM and total matter power spectra in massive neutrino cosmologies, where either the value of $\sigma_{8,cc}$ or $\sigma_{8,mm}$ are matched to the $\Lambda$CDM ones. We observe that on nonlinear scales, $0.2<k<1\,\kMpc$, a $\Lambda$CDM model with a given $\sigma_8$ is nearly indistinguishable from a massive neutrino model with the same value for $\sigma_{8,cc}$. A lower agreement is found when the match is done in terms of $\sigma_{8,mm}$, \ie in terms of the normalisation of the linear {\em total matter} power spectrum, since the relevant nonlinear evolution is given by CDM perturbations, and the amplitude normalisation can not take correctly into account the contribution from remaining linear neutrino perturbations (this should be taken into consideration especially in weak-lensing analyses, which probe directly the total matter power spectrum). This result represents the well known degeneracy between \smnu\ and $\sigma_8$ at small scales, and implies that the enhanced, nonlinear suppression of the matter power spectrum on nonlinear scales can be obtained tuning the normalisation of the linear one, without resorting to peculiar effects of massive neutrinos. Nonetheless, here we stress that, such kind of degeneracy can be broken when observing the power spectra on a much larger range of scales, $0.01<k<5\,\kMpc$, by means of future large sky galaxy surveys as, \eg, Euclid.

\section{Halos mass function}
\label{sec:mf}

The abundance of massive clusters as a function of redshift is a key cosmological probe for  neutrino mass measurements \cite{WangEtal2005, MantzAllenRapetti2010, CarboneEtal2012}. In particular, massive neutrinos could alleviate the tension between the value of $\sigma_8$ inferred from the primary CMB anisotropies \cite{Planck2013parameters,Planck2015parameters} and the one obtained from low redshift probes \cite{WymanEtal2013, BattyeMoss2014, CostanziEtal2014, BattyeCharnockMoss2014, LeistedtPeirisVerde2014, Planck2015szhalos, MantzEtal2015, RoncarelliCarboneMoscardini2015}. A careful calibration of the halo mass function in simulations with a massive neutrino component is, therefore, required to avoid systematic effects in the determination of cosmological parameters. 

The overall larger volume and higher mass resolution of the DEMNUni simulations with respect to previous studies (see \fig{fig:sims}) allow an improved analysis of the abundance of massive clusters over several decades in mass. In this work, we identify halos in two different ways: using a Friends-of-Friends (FoF) algorithm with linking length equal to $0.2$ times the mean inter-particle distance, and with the \textsc{SUBFIND-SO} code included in the \gadget\ package, which identifies halos whose spherical overdensity is $200$ times the mean background matter density, $M_{200} = (4 \pi / 3)\, R_{200}^3\, 200\, \bar{\rho}$, as already mentioned in \S\ref{sec:sim}. FoF halo masses have been corrected following the empirical prescription by \cite{WarrenEtal2006}, and, for the present analysis, no halos with less than $50$ particles have been considered.

The halo mass function, \ie the number density $n(M)$ of halos of mass between $M$ and $M+dM$, is often expressed as
\be
\label{eq:f}
n(M) = \frac{\bar{\rho}_c}{M}\,f(\sigma,z) \frac{d \ln\sigma^{-1}}{d M}\,dM \,, 
\ee
where most of the cosmological information is encoded in the variance of the matter distribution in the linear regime
\be
\label{eq:sigma}
\sigma^2(R,z)=\int d^3k\, P(k,z)\,W_R^2(k)\,,
\ee 
smoothed on the scale $R$ with a Top-Hat filter in real-space $W_R(k)$\footnote{To partially remove volume effects we use the fundamental frequency of the box, $k_f = 2 \pi/2000\kMpc$, as the lower integration limit in \eq{eq:sigma}. }. The function $f(\sigma,z)$ can be either predicted in the Press-Schechter framework \cite{PressSchechter1974} (see \cite{MussoSheth2012, ParanjapeShethDesjacques2013}  for recent results) or fitted to numerical simulations (see, \eg\,\cite{TinkerEtal2008, CrocceEtal2010, WatsonEtal2013}). The smoothing scale and the halo mass $M$ are related by the choice of the filter $W(kR)$, and, in the Top-Hat case, is given by 
\be
\label{eq:M-R}
M = \frac{4\pi}{3}\,\bar{\rho}_{c}\,R^3 \,.
\ee
Let us notice that in \eq{eq:f} as in \eq{eq:M-R}, in the case of massive neutrino models, we need to use the background density of the cold (rather than total) matter component $\bar{\rho}_c$ to define halo masses, since, as largely shown in the literature, the contribution of neutrinos (both bounded and unbounded) to the mass of CDM halos is completely negligible \cite{BrandbygeEtal2010,MarulliEtal2011,IchikiTakada2012}. 

Another potential ambiguity in massive neutrino scenarios is that the variance $\sigma(R,z)$ in \eq{eq:f} could in principle be computed for cold or total matter perturbations. However, as shown in \cite{IchikiTakada2012,CastorinaEtal2014}, the halo mass function is well described by analytic predictions, {\em and} by common fitting functions developed in the context of $\Lambda$CDM cosmologies, only if $\sigma=\sigma_{cc}$, \ie the variance of the matter distribution is computed from the cold matter linear power spectrum $P_{cc}(k)$. Here we summarise the physical mechanism behind this choice, referring the reader to \cite{IchikiTakada2012, CastorinaEtal2014, LoVerdeZaldarriaga2014, LoVerde2014B} for further details. In the spherical collapse model, assuming general relativity, the evolution of a region is controlled by the amount of mass inside its initial volume. For a Gaussian random field, as the linear density field $\delta$, a natural choice to describe the system is therefore its variance smoothed on the scale $R$ associated to the initial region, $\sigma(R)$.  Due to the tiny value of neutrino masses, the scale $R$ is much smaller than the free streaming length of massive neutrinos $\lambda_{\rm FS}$, typically tens of Megaparsec, and therefore neutrino perturbations are vanishingly small inside the collapsing region. This implies that $\sigma(R)$ should be computed using the CDM plus baryon perturbations only. A similar argument applies to the critical overdensity required for collapse at a given redshift, $\delta_{cr}$, that depends on neutrino masses only through their effect on the background evolution \cite{IchikiTakada2012}. Note that $\sigma_{cc}\ge\sigma_{mm}$ always (see \fig{fig:PL}) implying that for a given cosmology, using $\sigma_{cc}$ in the computation of the mass function leads to predicting more halos than using $\sigma_{mm}$ (see \cite{CostanziEtal2013B} for implications in galaxy cluster counts observations). Of particular importance is that, as shown in \cite{CastorinaEtal2014}, the universality of the function $f(\sigma,z)$  with respect to different cosmologies is recovered only for $\sigma=\sigma_{cc}$.
\begin{figure*}[t]
\begin{center}
\setlength{\tabcolsep}{0.01pt}
\begin{tabular}{c c}
\includegraphics[width=0.5\textwidth]{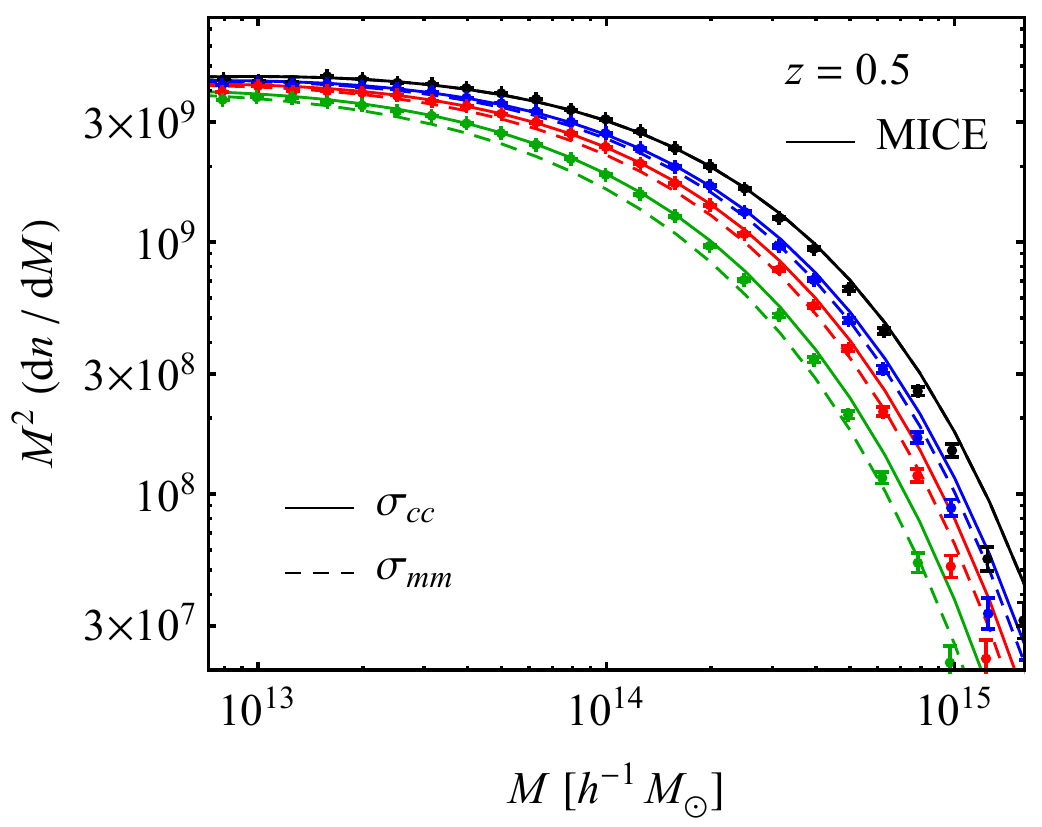}&
\includegraphics[width=0.5\textwidth]{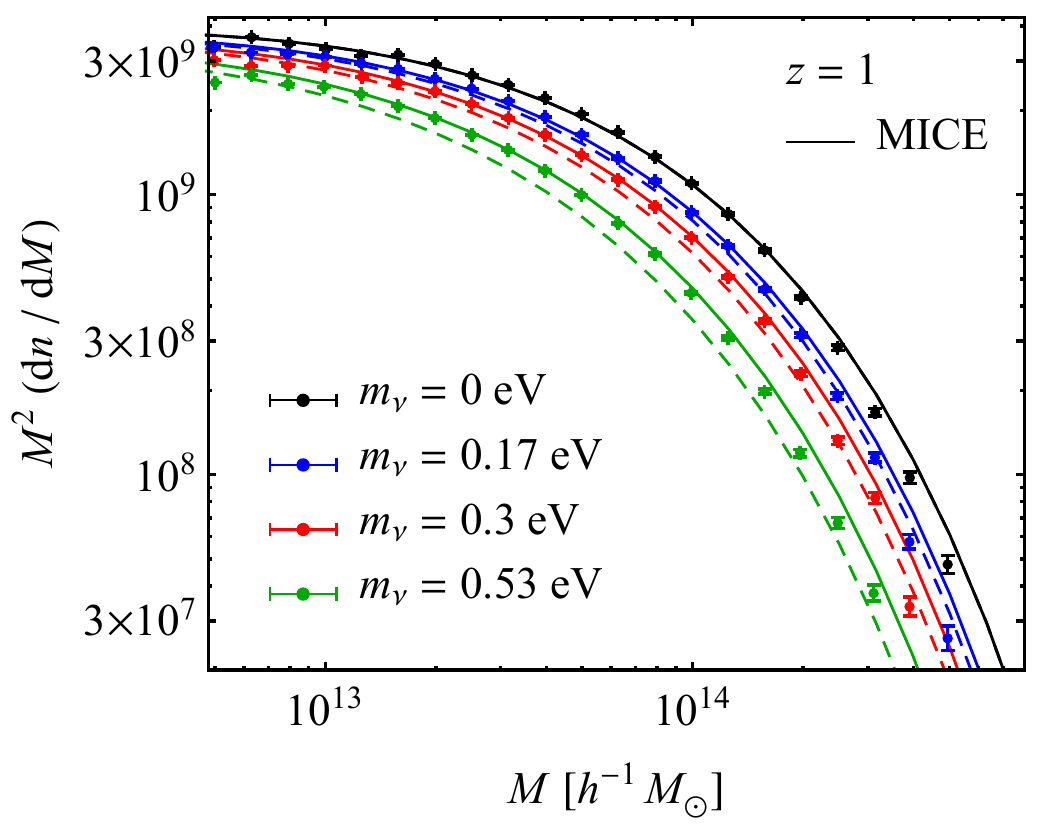}
\end{tabular}
\end{center}
\caption{\label{fig:HMF_FoF} 
Massive neutrino effects on the mass function of FoF halos. Data points show the mass function measured from the DEMNUni simulations for \smnut. Errors are derived from the assumption of a Poisson distribution for each bin. Theoretical predictions are obtained in terms of the fitting formula of \cite{CrocceEtal2010} as a function of $\sigma_{cc}(M)$ ({\em continuous curves}) and $\sigma_{mm}(M)$ ({\em dashed curves}). Different panels correspond to redshifts $z=0.5,\,1$.}
\end{figure*}
\begin{figure*}[t]
\begin{center}
\setlength{\tabcolsep}{0.01pt}
\begin{tabular}{c }
\includegraphics[width=1.0\textwidth]{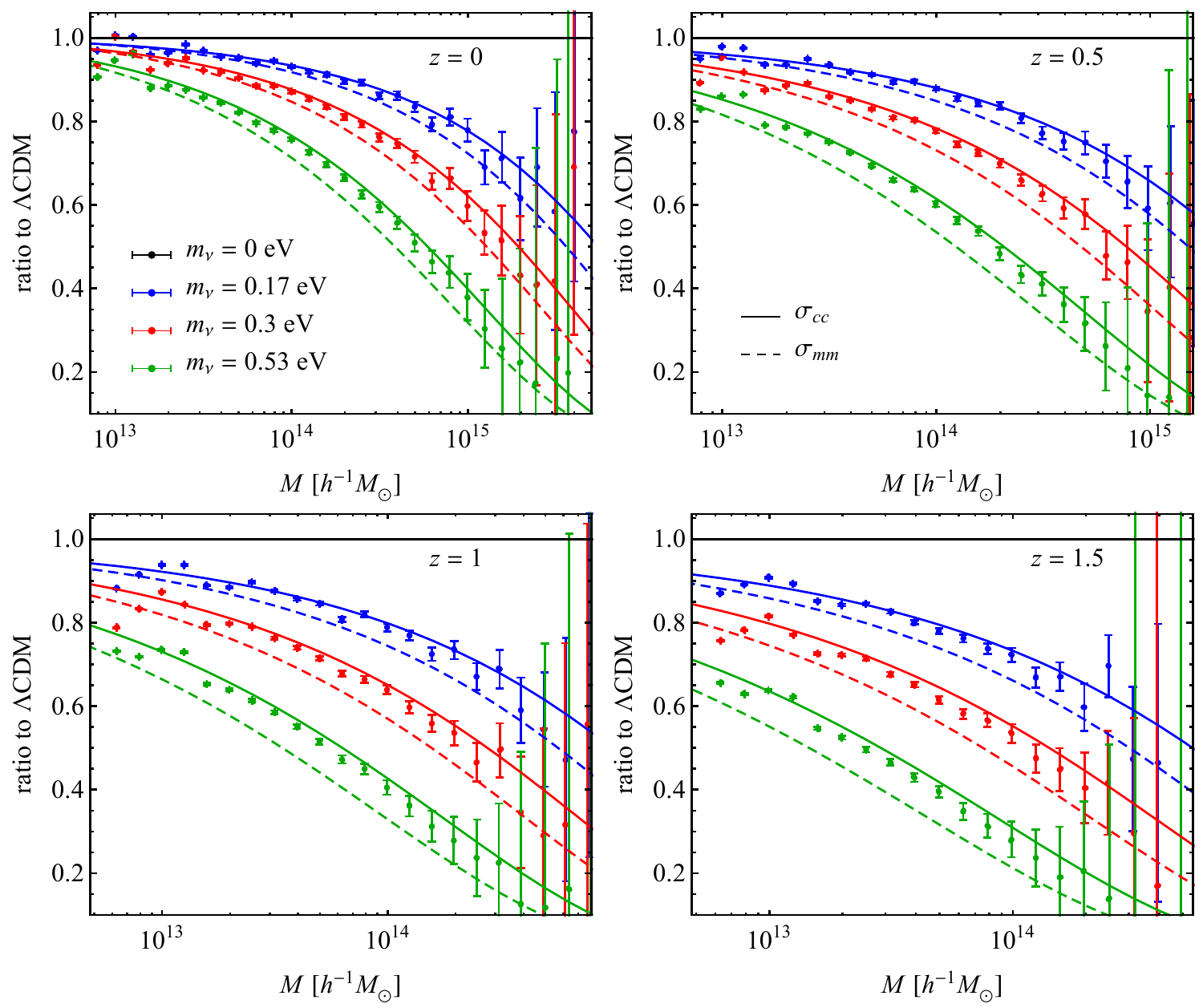}
\end{tabular}
\end{center} 
\caption{\label{fig:HMF_FoF_ratios} 
Massive neutrino effects on the mass function of FoF halos. Data points show the ratio of the mass function measured from the DEMNUni suite for  \smnu$=0.17,\,0.3,\,0.53$ to the massless case. Errors are derived from the assumption of a Poisson distribution for each bin. Theoretical predictions are obtained in terms of the fitting formula of \cite{CrocceEtal2010}  as a function of $\sigma_{cc}(M)$ ({\em continuous curves}) and $\sigma_{mm}(M)$ ({\em dashed curves}). Different panels correspond to redshifts $z=0,\,0.5,\,1,\,1.5$ ({\em clockwise from the top-right panel}).}
\end{figure*}

In \fig{fig:HMF_FoF} we show the suppression of the FoF halo abundance due to free streaming massive neutrinos. The points show measurements from the DEMNUni simulations, at two redshifts $z=0.5,\,1$, while the theoretical predictions are obtained in terms of the fitting formula of \cite{CrocceEtal2010} as a function of $\sigma_{cc}(M)$ ({\em continuous curves}) and $\sigma_{mm}(M)$ ({\em dashed curves}). Let us notice that the volume and mass resolution of the DEMNUni simulations allow to recover the mass function of the FoF halos on a quite large range of scales, $5\times 10^{12} \lesssim M \lesssim 10^{15}$ \msun. 

As a further inspection, in \fig{fig:HMF_FoF_ratios} we show the ratios, over the $\Lambda$CDM case, of the FoF halo mass function measured from the DEMNUni simulations at four different redshifts, $z=0,\,0.5,\,1,\,1.5$. Again, we compare these measurements to the predictions from the MICE fit to FoF halos in \cite{CrocceEtal2010}. Note that the MICE fit has been obtained using simulations of $\Lambda$CDM cosmologies alone, and it is non-universal in redshift, \ie $f(\sigma,z)$ does present an explicit redshift dependence on top of the $\sigma(z)$ one.

For the massive neutrino cosmologies we find that the cold matter prescription, with $\bar{\rho}=\bar{\rho}_c$ and $\sigma=\sigma_{cc}$ in \eq{eq:f}, reproduces very well the measurements when the MICE formula is assumed for the function $f(\sigma,z)$. In fact, we find the same level of agreement encountered for the $\Lambda$CDM model, therefore showing the same level of non-universality with respect to redshift. At high masses the measurements in the simulations drop below the MICE fit, because of the smaller volume of the DEMNUni simulations with respect to the MICE simulations. Our mass function measurements largely confirm the findings of \cite{CastorinaEtal2014}, but we stress that the DEMNUni simulations allow to extend such analysis by a factor of 8 in resolution while accounting, due to their larger volume, for the large-scale modes particularly relevant for this kind of studies \cite{CrocceEtal2010}.
\begin{figure*}[t]
\begin{center}
\setlength{\tabcolsep}{0.01pt}
\begin{tabular}{c }
\includegraphics[width=1.0\textwidth]{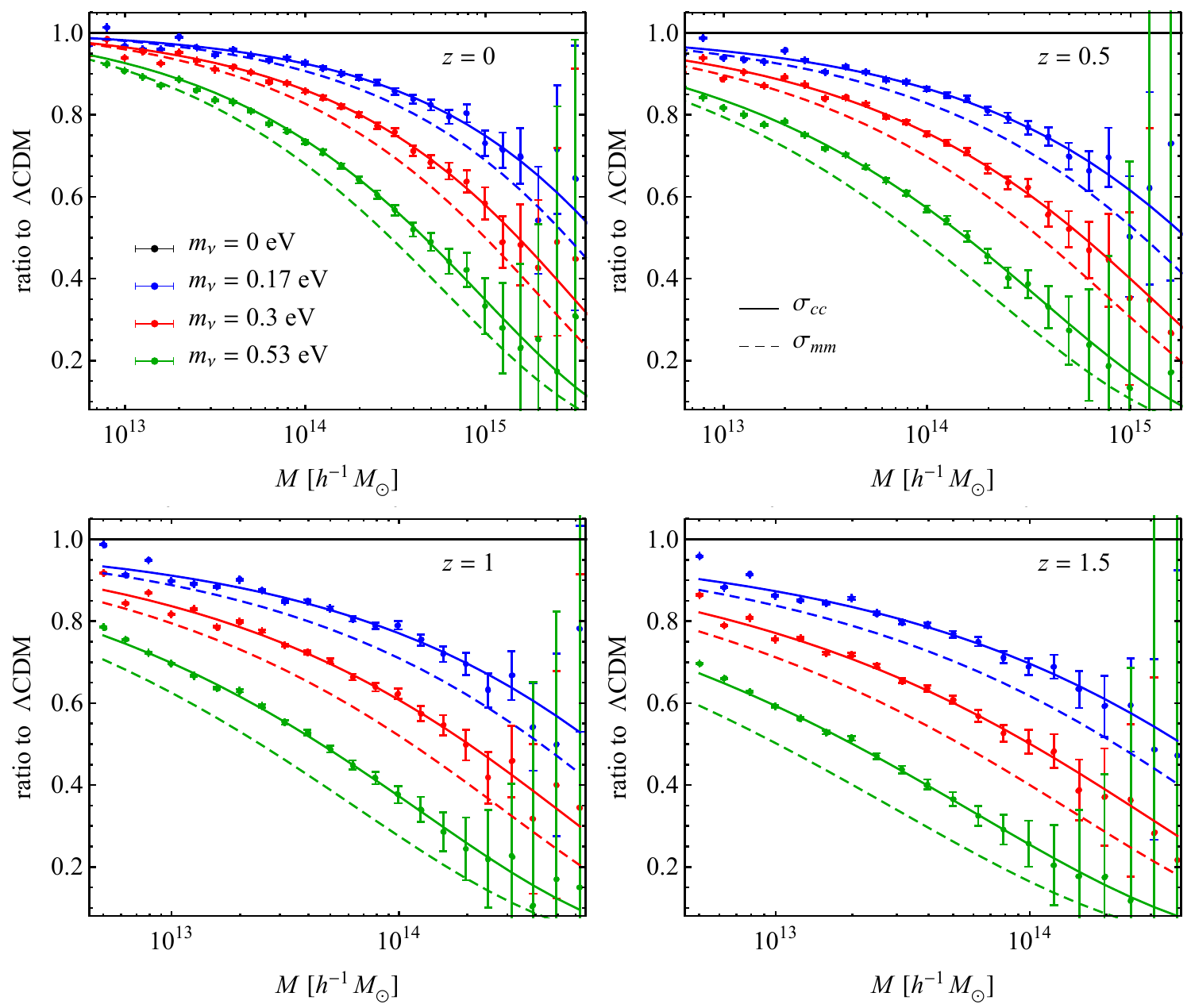}
\end{tabular}
\end{center} 
\caption{\label{fig:HMF_SO} 
Same as \fig{fig:HMF_FoF} but for Spherical Overdensity halos. Theoretical predictions, using $\sigma_{cc}$ ({\em continuous curves}) and using $\sigma_{mm}$ ({\em dashed curves}), have been obtained using the formula of \cite{TinkerEtal2008}.}
\end{figure*}

We repeat the analysis for Spherical Overdensity (SO) halos, identifying halos as spherical regions with a mean matter density equal to two hundred times the background density, as described in \S\ref{sec:sim}. The measurements, shown as ratios to the $\Lambda$CDM case, are presented in \fig{fig:HMF_SO}, where now we use the Tinker {\em et al.} fitting formula as a reference for SO halos \cite{TinkerEtal2008}. Also in this case we observe that the Tinker {\em et al.} fit, developed in the $\Lambda$CDM framework, provides a good fit to simulations with a massive neutrino component when $\bar{\rho}=\bar{\rho}_c$ and $\sigma=\sigma_{cc}$ are used instead of their total dark matter counterparts.
Finally, we can assert that a parametrisation of the halo mass function in terms of the CDM power spectrum gives a much better description of measurements from simulations, including massive neutrinos, with respect to the DM power spectrum, both for FoF and SO halos, therefore confirming the results of \cite{IchikiTakada2012,CastorinaEtal2014,CostanziEtal2013}.

\section{Halo bias and redshift space distortions}
\label{sec:halops}

Measurements of the linear bias of FoF halos in massive neutrinos simulations have been presented already in \cite{CastorinaEtal2014}, where the authors showed how halo bias at large-scales is scale-independent, as in standard $\Lambda$CDM cosmologies, only if defined with respect to the {\em cold}, rather than {\em total}, matter perturbations. It follows that the halo bias defined with respect to the total matter fluctuations presents a spurious scale-dependence simply arising from the difference between the cold and total matter power spectra, see Fig.\,\ref{fig:PL}. 

In this section we check, in the first place, these previous results taking advantage of the higher mass resolution of the DEMNUni simulations, with the aim of analysing halo population of sensibly lower mass. In addition, we perform a preliminary analysis of Redshift Space Distortions (RSD) in massive neutrino models with a specific attention to possible consequences of the aforementioned ambiguity in the halo bias definition. More specifically, we test the Kaiser limit \cite{Kaiser1987} of RSD, as the simplest, observable feature of anisotropic galaxy clustering directly dependent on the proper definition of bias. The main goal is to highlight possible systematic effects in the determination of the growth rate, which, in the presence of massive neutrinos, is not simply a function of time, as in $\Lambda$CDM models, but presents a mild scale-dependence as well. This analysis is therefore rather complementary to the one of \cite{MarulliEtal2011} where systematic errors on the determination of RSD parameters induced by neglecting neutrino masses have also been studied.

While we have analysed, for this purpose, three distinct halo populations characterised by distinct mass thresholds, here we focus on the lower one, corresponding to $M\ge 10^{13}\Ms$, as less affected by shot-noise and characterised by smaller nonlinear bias corrections. We consider, however, this threshold at different redshifts, presenting the measurements at $z=0.5$ and $z=1$ as the most relevant for future observations.  

\subsection{The halo power spectrum in real space}
\label{ssec:real}

In the local bias approach \cite{FryGaztanaga1993}, one can expand the halo overdensity $\delta_h$ as a power series in $\delta_m$, such that on very large scales the bias relation  can be written as
\be\label{eq:bias}
\delta_h \simeq b\, \delta_m\,,
\ee
where the constant $b$ is the linear bias factor for the halo population. 
From the equation above (and from the definition of power spectrum) it follows, in $\Lambda$CDM cosmologies, a linear relation between the matter and halo power spectra given by
\be
P_{hh}(k)\simeq b^2\,P_{mm}(k)\,.
 \ee
Linear bias can therefore be measured in N-body simulations as $b^2\equiv P_{hh}/P_{mm}$ in terms of the measured power spectra $P_{hh}$ and $P_{mm}$. In massive neutrino cosmologies, however, the $P(k)$ at the RHS of the above expression can be either the cold or total matter power spectrum, leading to two possible definitions of linear halo (and galaxy) bias, that is
\be\label{eq:biasc}
b_{c} = \sqrt \frac{P_{hh}}{P_{cc}}
\ee
or
\be\label{eq:biasm}
\qquad b_{m} = \sqrt \frac{P_{hh}}{P_{mm}}\,.
\ee
Analogous definitions can be given in terms of the cross halo-matter power spectra, $b_{c,\times}=P_{hc}/P_{cc}$ and $b_{m,\times}=P_{hm}/P_{mm}$. The results from \S\ref{sec:mf} and an analogy with $\Lambda$CDM cosmologies suggest that bias coefficients are scale-independent only if defined with respect to the cold matter field. 

\begin{figure}[t]
\begin{center}
\includegraphics[width=.98\textwidth]{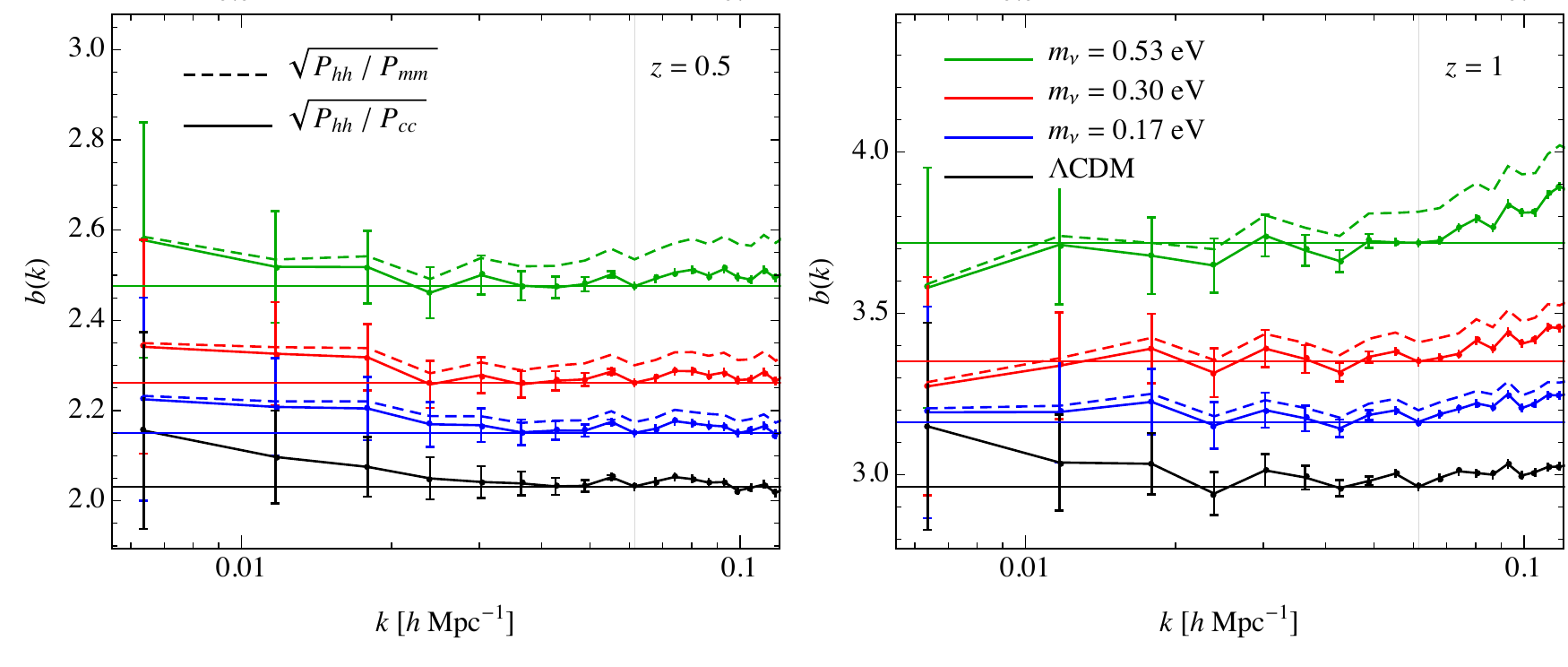}
\caption{\label{fig:bias} 
Linear halo bias from the halo auto-power spectrum as a function of scale for halos of mass $M>10^{13}\Ms$ in the DEMNUni simulations at redshift $z=0.5$ ({\em left panel}) and $z=1$ ({\em right panel}). Continuous curves with error bars show the bias defined w.r.t. the cold matter power spectrum as $b(k)\equiv\sqrt{P_{hh}(k)/P_{cc}(k)}$, while dashed curves denote the bias defined w.r.t. the total matter perturbations, $b(k)\equiv\sqrt{P_{hh}(k)/P_{mm}(k)}$. Horizontal lines show the value of the linear bias at $k\simeq 0.06\kMpc$, assumed here, approximately, as the limit of validity for linear theory.}
\end{center}
\end{figure}

In \fig{fig:bias} we present the measurements of halo bias on linear scales from the DEMNUni simulations for halos of masses $M\ge 10^{13}\Ms$ at redshifts $z=0.5,\,1$. We consider the bias determined according to both \eq{eq:biasc} and \eq{eq:biasm}. In all the cases, the bias defined with respect to the total matter perturbations exhibits a larger scale-dependence than the one defined w.r.t. the cold matter component, confirming the findings in \cite{CastorinaEtal2014}. In the rest of the section we will explore possible implications of this fact for the description of RSD at large scales.

\subsection{The halo power spectrum in redshift-space}
\label{ssec:rsd}

The biasing between the galaxy and matter distributions is not the only effect to be taken into account for a correct estimate of the matter power spectrum from galaxy redshift observations. In a real survey, the proper motions of galaxies with respect to the homogeneous expansion of the Universe affect the determination of their distance along the line-of-sight (see, \eg\, \cite{Kaiser1987, Scoccimarro2004, ReidWhite2011, BianchiChiesaGuzzo2015}). However, on sufficiently large scales these redshift space distortions provide information on the peculiar velocity field of matter perturbations, in particular on the growth rate of the density field, and, as a result, they are extensively used to constrain cosmological parameters and test deviation from standard gravity \cite{GuzzoEtal2008, DelaTorreEtal2013, ContrerasEtal2013, BeutlerEtal2014A}.  In preparation for future large spectroscopic surveys, the effect of neutrino masses on RSD modelling needs to be carefully investigated to avoid fake signatures of Modified Gravity. In the following we present a preliminary assessment of the scale-dependence of the growth rate induced by the free streaming length of neutrinos, and its measurement from the DEMNUni simulations.

In the standard cosmological model, and in the large-scale limit where linear theory applies, the distortion induced by peculiar velocities on the density contrast $\delta_m$ can be written in Fourier Space as \cite{Kaiser1987}
\be
\label{eq:kaiserM}
\delta_{m,s}(\vec{k}) = (1+ f \mu^2)\,\delta_m(k)\,,
\ee
where $\mu= \vec{k} \cdot \hat z / k$ is the angle between the line-of-sight and the wave vector $\vec k$, while $f(a)$ is the linear growth rate, defined as the logarithmic derivative of the linear growth factor $D(a)$, that is
\be
f(a) \equiv \frac {d  \ln D(a)}{d \ln a}\,.
\ee  
A similar expression holds for linearly biased tracers which follow the matter flow (\ie with no velocity bias\footnote{Note that the assumption of no velocity bias is better justified for halos w.r.t. the cold matter perturbations rather than w.r.t. the total matter ones.}). For instance, the halo overdensity in redshift space, $\delta_{h,s}$, can be written as
\be
\label{eq:kaiserH}
\delta_{h,s}(\vec{k}) = (b + f \mu^2 ) \delta_m \equiv (1+ \beta \mu^2) \,b \,\delta_m (k)\,,
\ee
where $b$ is the linear scale-independent bias, and we define $\beta \equiv f/b$.  It follows that the halo power spectrum in redshift space can be written as the product of a polynomial in $\mu$ times the halo power spectrum in real space. Since the halo power spectrum in real space depends only on the modulus of the wave vector, the angular dependence induced by RSD is entirely encoded in the pre-factor of \eq{eq:kaiserH}.  As a result, the redshift-space halo power spectrum can be decomposed in multipoles using just the first three even Legendre polynomials $L_\ell(\mu)$
\be
P_{hh,s}(\vec{k}) = (1+ \beta \mu^2)^2\, P_{hh}(k) = \sum_{l=0,2,4} P_{hh,\ell} \,L_{\ell}(\mu) \,.
\ee
The monopole, quadrupole and hexadecapole coefficients read
\bea
\label{eq:P0}
P_{hh,0} (k) &=& \left(1 + \frac{2}{3} \beta + \frac{1}{5} \beta^2 \right) P_{hh} (k) \\
P_{hh,2} (k) &=& \left(\frac{4}{3} \beta + \frac{4}{7} \beta^2 \right) P_{hh} (k) \\
P_{hh,4} (k) &=&  \frac{8}{35} \beta^2 P_{hh} (k),
\eea
where the RSD parameter $\beta$ determines the relative amplitude of the multipoles. In a $\Lambda$CDM cosmology, $\beta$ is predicted to be scale-independent on large scales, as a direct consequence of the scale-independence of both linear bias and growth rate. On the other hand, both quantities might be scale-dependent when massive neutrinos are present.

\begin{figure}[t]
\begin{center}
\includegraphics[width=1\textwidth]{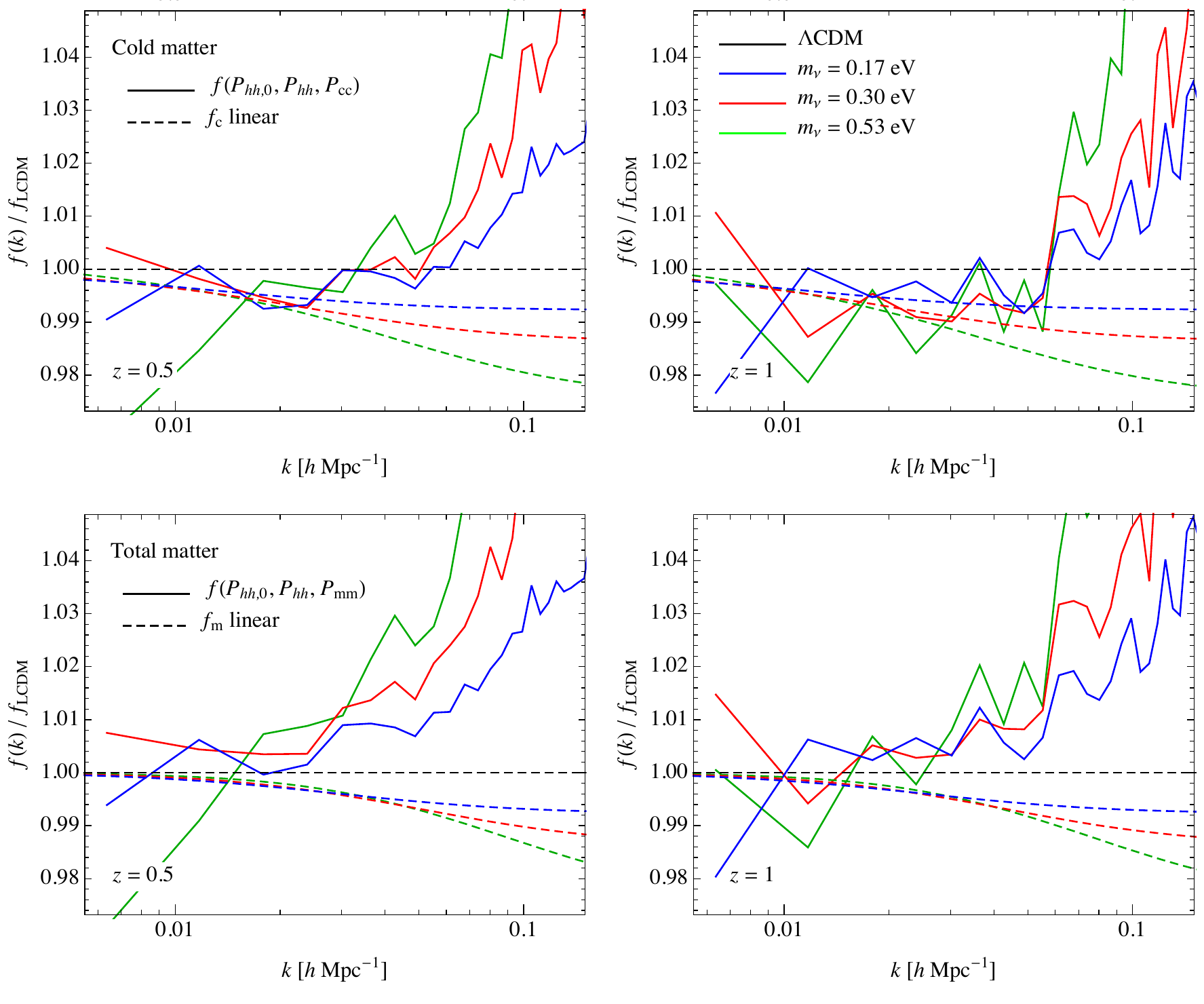}
\caption{\label{fig:grate}
{\em Top panels}: Linear growth rate obtained from measurements of the redshift-space (monopole) and real-space halo power spectrum according to \eq{eq:P0} and from measurements of the linear halo bias $b(k)=\sqrt{P_{hh}(k)/P_{cc}(k)}$ at redshifts $z=0.5$ ({\em left}) and $z=1$ ({\em right}). Dashed curves denote the linear theory (scale-dependent) predictions for cold matter perturbations, $f_c(k)$. All the quantities are shown in terms of their ratio to the corresponding $\Lambda$CDM values. {\em Bottom panels}: Same as in the top panels, but assuming the linear halo bias defined in terms of total matter perturbations. In this case, the comparison is made to the linear theory prediction for the total matter growth rate, $f_m(k)$.}
\end{center}
\end{figure}

In the first place, in massive neutrino scenarios, the growth rate of cold matter perturbations, $f_c$, becomes scale-dependent, as a natural consequence of the scale-dependent growth function $D_c$. In particular, the {\em small-scale} asymptotic suppression expected in linear theory is given by \cite{BondEfstathiouSilk1980}
\be
\frac{f_c(k)}{f_{\Lambda{\rm CDM}}}\stackrel{k\gg k_{\rm FS}}{\longrightarrow} \frac{1}{4}\left(5-\sqrt{25-24\,f_\nu}\right)\simeq 1-\frac35 \,f_\nu\,,
\label{eq:fc_small}
\ee
and corresponds to a 2.4\% effect for the largest value of the neutrino fraction assumed here, that is $f_\nu\simeq 0.039$ in the \smnu$=0.53$ eV model.  However, for the same model, the suppression is below the percent level on large scales, $k\lesssim 0.03\kMpc$. In the case of the total matter growth rate, $f_m(k)$, the suppression, again for the \smnu$=0.53$ eV model, reaches the 1\% level at slightly smaller scales, $k\simeq 0.05\kMpc$ (linear theory predictions for both the ratio $f_c/f_{\Lambda{\rm CDM}}$ and $f_m/f_{\Lambda{\rm CDM}}$ are shown as dashed curves in \fig{fig:grate}).

In the second place, if we defined the halo bias w.r.t. the total matter field, according to \eq{eq:biasm}, we would add to the quantity $\beta$ an additional source of scale-dependence. Let us notice here that, while one would consider reasonable that the choice of $b_c$, \eq{eq:biasc}, is consisted with the choice of $f=f_c$ in the definition of $\beta=f/b$, we cannot exclude {\em a priori} other possibilities (including, for instance, a mixed $\beta=f_m/b_c$), since the physical origins of the two contributions in \eq{eq:kaiserH} are in fact distinct. However, in what follows, we will limit ourselves to compare the two cases $\beta=f_c/b_c$ and $\beta=f_m/b_m$, the latter corresponding to na\"ively neglecting any distinctions between matter contributions. 
\begin{figure}[t]
\begin{center}
\includegraphics[width=1\textwidth]{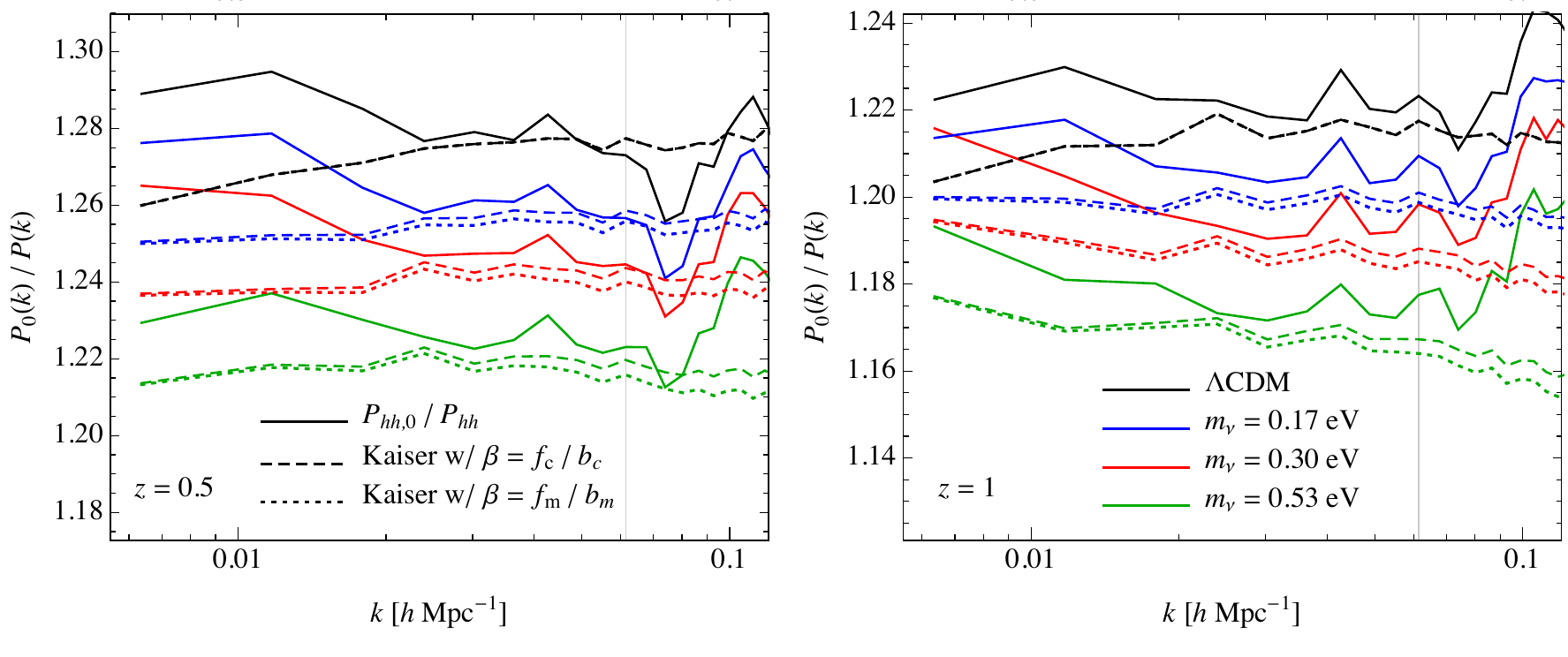}
\caption{\label{fig:monopole} 
Comparison of the ratio between the monopole of the redshift-space halo power spectrum to the real-space halo power spectrum, $P_{hh,0}(k)/P_{hh}(k)$ ({\em continuous curves}) against the value predicted by the Kaiser formula, \eq{eq:P0}, as a function of $\beta=f_c/b_c$ ({\em dashed curves}) or $\beta=f_m/b_m$ ({\em dotted curves}). $f_c(k)$ and $f_m(k)$ correspond to the linear theory growth rate respectively of cold and total matter perturbations, while the $b_c$ and $b_m$ are the {\em measured} values of the halo bias according to the two definitions of \eqs{eq:biasc} and (\ref{eq:biasm}). Notice that we {\em do not} use best fit values for the bias parameters, but we use instead the measured $b(k)$ as a function of scale.}
\end{center}
\end{figure}

In all the cases, we expect the Kaiser factor $\beta$ to exhibit a scale-dependence, which, by choosing $\beta=f_c/b_c$, will result only from the growth rate  $f_c(k)$. In the other case, $\beta=f_m/b_m$, a partial compensation occurs between the linear bias, which increases with the wave-number, and the growth factor that actually decreases with $k$.  In \fig{fig:monopole} we show the ratio of the monopole of the redshift-space halo power spectrum, $P_{hh,0}(k)$, to the halo power spectrum in real space, $P_{hh}(k)$, for the models listed in Table \ref{tab:par}, and we compare the measurements from the DEMNUni simulations to the corresponding predictions provided by the Kaiser formula written both in terms of $\beta=f_c/b_c$ and $\beta=f_m/b_m$. In such predictions we {\em do not} use best fit values for the bias parameters, but we use instead the {\em measured} $b(k)$ where the scale-dependence is affected by cosmic variance. While the noisy measurements do not allow to clearly determine the scale-independence of such a quantity, we notice that $\beta=f_c/b_c$ does provide a slightly better prediction than $\beta=f_m/b_m$, as compared to simulations.

As a further test, in \fig{fig:grate}  we consider directly the growth rate $f$ obtained from measurements of the monopole of the redshift-space halo power spectrum, $P_{hh,0}(k)$, and the real-space halo and cold matter power spectra, $P_{hh}(k)$ and $P_{cc}(k)$ respectively, under the assumption that the Kaiser formula provides a good description of RSD at large scales. Assuming $\beta=f/b_c$, from \eq{eq:P0} we have
\be\label{eq:fmes}
f(k)=\sqrt{\frac{P_{hh}(k)}{P_{cc}(k)}}\,\frac{1}{3}\,\left[\sqrt{45\,\frac{P_{hh,0}(k)}{P_{hh}(k)}-20}-5\right]\,
\ee
while assuming $\beta=f/b_m$ we obtain a similar expression where $P_{mm}$ replaces $P_{cc}$.  In order to further reduce cosmic variance, we show the recovered value of $f$ as ratio to the $\Lambda$CDM case. These measurements, as a function of $k$, can then be directly compared with linear theory predictions for $f_c(k)$ and $f_m(k)$. 

The top panels of \fig{fig:grate} show $f$ obtained from \eq{eq:fmes} under the cold matter hypothesis compared to the predictions for $f_c(k)$ at redshift $z=0.5,\,1$. We notice that, in the higher redshift case in particular, despite the noise, the measurements are consistent with the linear predictions at large scales when $f$ is obtained from the measured $P_{cc}$. A greater discrepancy instead is observed when $f$ is obtained from the measured total matter $P_{mm}(k)$, as compared to the predictions for $f_m(k)$ in the bottom panels. If such discrepancy will be confirmed by future investigations, we could expect that using $b_m$ as a definition for bias in the Kaiser formula could lead to a systematic error on the determination of the growth rate at the level of 1-2\%.

Clearly this represents a very preliminary analysis of possible systematic effects in the determination of the growth rate $f(k)$ in the context of massive neutrinos cosmologies. If the description of halo clustering in terms of cold matter perturbation does indeed represents the correct approach, this test can serve as a confirmation that the DEMNUni simulations reproduce linear theory predictions for the growth rate including, to a certain extent, its scale-dependence. We reserve for future work a more detailed analysis of RSD effects on the matter density field as well as on realistic mock galaxy distributions, in massive neutrino scenarios. Of particular interest would be any description of nonlinearity that could extend theoretical predictions to smaller scales.

\section{Conclusions}
\label{sec:conclusions}

In this work we have presented an analysis of the LSS clustering features extracted from the DEMNUni simulation suite, a set of large-volume and high-resolution N-body simulations that include massive neutrinos as a particle component. The volume and resolution of these simulations allow the test of several cosmological observables, from the galaxy power spectrum to weak lensing statistics and CMB secondary anisotropies \cite{CarbonePetkovaDolag2015}. In this respect, the DEMNUni simulations represent the largest effort to date to include the effects of neutrinos mass in numerical predictions of the large-scale structure. As such, they serve, in the first place, the purpose of investigating our ability to constrain the properties of such important particle for both the standard cosmological model as the standard model of particle physics. At the same time, they can help assessing the effects of massive neutrino on the expected accuracy {\em and} precision in the determination of other cosmological parameters, first and foremost those describing possible departures from the $\Lambda$CDM paradigm.

We have analysed basic quantities such as the matter power spectrum, the halo mass function, and halo clustering both in real and redshift space, comparing them with the simplest and most common theoretical predictions. This analysis is intended as a first step toward a more accurate test of the latter, taking advantage, however, of recent developments in the understanding of halo formation and clustering.

The most fundamental quantity, in this respect, is the matter power spectrum. In Section~\ref{sec:matterps}
we presented measurements of the power spectrum of each individual component of the total matter density, that is the cold dark matter and baryons (here treated on the same footing and called CDM) and the neutrinos, along with their cross-power spectrum and the total matter power spectrum, given by the weighted sum of the three. The analysis in terms of distinct components allows to test early hypothesis regarding the possibility of neglecting the nonlinear evolution of neutrino perturbations in analytical predictions \cite{SaitoTakadaTaruya2008}. In this respect we point-out, with direct reference to our measurements, that the neutrino power spectrum, as well as the cross-power spectrum provide a significant contribution to the total matter power spectrum, \eq{eq:Pmm}, only at large-scales, where the linear approximation is sufficient. Since the total matter power spectrum, $P_{mm}(k)$, and the cold matter power spectrum, $P_{cc}(k)$, are the only quantities directly related to actual observables, a 1\% accuracy appears to be achievable even neglecting the nonlinear evolution of neutrino perturbations. With this in mind, we focused on the accurate description of nonlinearity of the cold matter component {\em alone}, considering various predictions in perturbation theory, and using, as input quantity, the simple cold matter linear power spectrum. While this is clearly an effective, not rigorous, approach, we have shown that the accuracy provided by PT techniques, and their limits of validity on the matter power spectrum in $\Lambda$CDM cosmologies, can be achieved as well for massive neutrino models in a rather simple way. In a completely similar fashion, fitting formulae obtained from $\Lambda$CDM simulations, and used beyond the perturbative regime like {\halofit}, can be safely applied to predictions of the nonlinear cold matter power spectrum, and simply extended, with the addition of the linear contributions from the neutrino auto- and CDM-neutrino cross-spectra, to the total matter power spectrum, retaining the same accuracy in presence of massive neutrinos as in a standard massless cosmology, without resorting to additional fitting parameters.

The analysis of the DEMNUni simulations has allowed us to confirm the results of \cite{CastorinaEtal2014, CostanziEtal2013} on halo abundances in massive neutrino cosmologies, \ie that the halo mass function is better described in terms of  the variance of cold matter fluctuations alone. Similarly, from the study of halo clustering, we have confirmed as well that a {\em constant} linear, halo bias is recovered in the large-scale limit only when the bias relation is considered between halo and cold (rather than total) matter distributions. Such peculiar ambiguity in the definition of halo (and galaxy) bias, motivated a preliminary test of the Kaiser formula for redshift-space distortions \cite{Kaiser1987}, where linear bias is a relevant parameter. We found, despite the yet large statistical uncertainty, some indications of the expected scale-dependence of the growth rate of matter perturbation, characteristic of massive neutrino cosmologies. If such indications are correct, then we have shown that defining the galaxy bias w.r.t. the cold matter component is necessary to avoid a systematic error on the determination of the growth rate $f(k)$ at the few percent level. We will return on the subject in more detail in future work.

\subsection*{Acknowledgements} 
We thank Jason Dossett, Carlo Schimd, Roman Scoccimarro for useful
discussions. We are especially grateful for the support of Margarita Petkova through the Computational Center for Particle and Astrophysics (C$^2$PAP). We are also particularly grateful to Francisco
Villaescusa-Navarro for making available to us the results of his
simulations used in Fig.~\ref{fig:cucchiaio}, and for a constant
exchange of ideas during the development of the results presented in
this paper. C.C. and K.D. thank Matteo Viel for providing the N-GenIC code
for initial conditions, modified to take into account a massive
neutrino particle component. The DEMNUni simulations were carried out
at the Tier-0 IBM BG/Q machine, Fermi, of the Centro Interuniversitario del Nord-Est per il Calcolo Elettronico (CINECA, Bologna, Italy), via the five million cpu-hrs budget provided by the Italian SuperComputing Resource Allocation (ISCRA) to the class--A proposal entitled ``The Dark-Energy and Massive-Neutrino Universe". C.C. is extremely grateful to CINECA system managers for the great help and support provided during the simulation production at Fermi, and the ``BigData" management and storage, via the new computing system PICO, which drives a large data repository shared among all the HPC systems at CINECA. C.C. thanks Federico Marulli, Lauro Moscardini, and Andrea Cimatti for useful discussions in the preparation of the DEMNUni proposal. C.C. acknowledges financial support from the INAF Fellowships Programme 2010. 
K.D. acknowledges the support by the DFG Cluster of Excellence ``Origin and Structure of the Universe". 
E.S. would like to express as well his gratitude to Ivan Girotto at ICTP for his invaluable kindness and technical help in the solution of issues related to the data analysis presented in this paper. E.S. acknowledges financial contribution from contracts ASI/INAF n.I/023/12/0 ‘Attivià relative alla fase B2/C per la missione Euclid. J.B. and C.C. also acknowledge financial support from the European Research Council through the Darklight Advanced Research Grant (n. 291521).

\bigskip

\appendix

%
%
%
%

\bibliographystyle{JHEPb}
\bibliography{Bibliography}

\end{document}